\documentclass[lettersize,journal]{IEEEtran}
\usepackage{graphicx}
\usepackage{amsmath,amsfonts}
\usepackage[colorlinks,
    linkcolor=blue,
    citecolor=blue,
    pdftex]{hyperref}
\usepackage[dvipsnames]{xcolor}
\usepackage{tabularx}
\usepackage{balance}
\usepackage{xspace}
\usepackage{mdframed}
\usepackage{booktabs}
\usepackage{hyperref}
\usepackage{listings}
\usepackage{color}
\usepackage{subcaption}
\usepackage{xcolor}
\usepackage{svg}
\usepackage{tcolorbox}
\usepackage{wrapfig}
\definecolor{codegreen}{rgb}{0,0.6,0}
\definecolor{codegray}{rgb}{0.5,0.5,0.5}
\definecolor{codepurple}{rgb}{0.58,0,0.82}
\definecolor{backcolour}{rgb}{0.95,0.95,0.92}

\definecolor{lightgray}{rgb}{.9,.9,.9}
\definecolor{darkgray}{rgb}{.4,.4,.4}
\definecolor{purple}{rgb}{0.65, 0.12, 0.82}

\lstdefinelanguage{JavaScript}{
  keywords={typeof, new, true, false, catch, function, return, null, catch, switch, var, if, in, while, do, else, case, break, let},
  keywordstyle=\color{magenta}\bfseries,
  ndkeywords={class, export, boolean, throw, implements, import, this,for},
  ndkeywordstyle=\color{magenta}\bfseries,
  identifierstyle=\color{black},
  sensitive=false,
  comment=[l]{//},
  morecomment=[s]{/*}{*/},
  commentstyle=\color{purple}\ttfamily,
  stringstyle=\color{red}\ttfamily,
  morestring=[b]',
  morestring=[b]"
}

\newtcolorbox[auto counter, number within=section]{promptbox}[2][]{
    enhanced,
    colback=white,       
    colframe=black,      
    colbacktitle=white,  
    coltitle=black,      
    boxrule=0.8pt,
    sharp corners,
    title={#2},
    label={#1},
    fonttitle=\bfseries,
    left=2mm,
    right=2mm,
    top=0mm,
    bottom=0mm,
}

\newcommand{\nboriginsinitial}{1,645,775\xspace}
\newcommand{\nborigins}{1,613,571\xspace}
\newcommand{\nbtopusers}{42\xspace}
\newcommand{\nbdead}{84,666\xspace}
\newcommand{\nbpfiverqthree}{17,875\xspace}

\newcommand{\nbrqtwosample}{154,832\xspace} 
\newcommand{\nbuniquecontributors}{96,581\xspace} 
\newcommand{\nbnonemptylocations}{22,349\xspace} 
\newcommand{\nbreposwithvalidlocation}{14,427\xspace}

\newcommand{\nbuniquecountries}{124\xspace}

\newcommand{\swh}{Software Heritage\xspace}
\newcommand{\pfive}{\texttt{p5.js}\xspace}
\newcommand{\supercollider}{\texttt{SuperCollider}\xspace}
\newcommand{\processing}{\texttt{Processing}\xspace}
\newcommand{\OF}{\texttt{openFrameworks}\xspace}
\newcommand{\nbreposbytopfortytwo}{17,156 \xspace}

\newcommand{\nbrqthreesample}{7,985\xspace}
\newcommand{\nbrawpfivesketches}{19,797\xspace}

\newcommand{\matproc}{{\small \texttt{Material and Processes}}\xspace}

\newcommand{\procaudio}{{\small \texttt{processed\_audio}}\xspace}
\newcommand{\proctext}{{\small \texttt{processed\_text}}\xspace}
\newcommand{\procimage}{{\small \texttt{processed\_image}}\xspace}
\newcommand{\synvisual}{{\small \texttt{synthesized\_visual}}\xspace}
\newcommand{\synaudio}{{\small \texttt{synthesized\_audio}}\xspace}
\newcommand{\syntext}{{\small \texttt{synthesized\_text}}\xspace}
\newcommand{\synimage}{{\small \texttt{synthesized\_image}}\xspace}
\newcommand{\randomness}{{\small \texttt{randomness}}\xspace}
\newcommand{\interaction}{{\small \texttt{Interaction}}\xspace}
\newcommand{\interactive}{{\small \texttt{interactive}}\xspace}
\newcommand{\noninteractive}{{\small \texttt{non-interactive}}\xspace}
\newcommand{\outcome}{{\small \texttt{Sensory Outcome}}\xspace}
\newcommand{\visual}{{\small \texttt{visual}}\xspace}
\newcommand{\auditory}{{\small \texttt{auditory}}\xspace}
\newcommand{\audiovisual}{{\small \texttt{auditory and visual}}\xspace}
\newcommand{\static}{{\small \texttt{static}}\xspace}
\newcommand{\timebased}{{\small \texttt{time\_based}}\xspace}

\DeclareRobustCommand{\cityicon}{\includegraphics[scale=0.52]{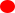}}
\DeclareRobustCommand{\areaicon}{\includegraphics[scale=0.52]{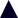}}

\title{Art in Humanity's Code}

\author{
    \IEEEauthorblockN{
        Benoit Baudry\IEEEauthorrefmark{1},
        Yogya Gamage\IEEEauthorrefmark{1},
        Nadia Gonzalez Fernandez\IEEEauthorrefmark{1},
        Lena MK\IEEEauthorrefmark{1},
        Roxana Pena Mendieta\IEEEauthorrefmark{1},
        Rafaela Souza Pinter\IEEEauthorrefmark{1},
        Stefano Zacchiroli\IEEEauthorrefmark{2}
    } \\
    \IEEEauthorblockA{\IEEEauthorrefmark{1}Université de Montréal, Montréal, Canada} \\
    \IEEEauthorblockA{\IEEEauthorrefmark{2} LTCI, Télécom Paris, Institut Polytechnique de Paris, Palaiseau, France} \\
    \IEEEauthorrefmark{1}{\{firstname.lastname\}@umontreal.ca}
    \IEEEauthorrefmark{2}{stefano.zacchiroli@telecom-paris.fr} 
}

\begin{document}

\maketitle
 
\begin{abstract}

Artist-led open-source libraries such as \processing or \OF have had a major impact on artists and designers who use code as a creative medium. In this work, we conduct the first large-scale empirical study of public code repositories that use these libraries. Our study dives into \nborigins code repositories collected from the \swh archive. Blending quantitative and qualitative methods, we investigate the diversity of practices and practitioners regarding code hosting, geographical distribution, and the characteristics and purposes of code-based creative works. 
Key findings include evidence of the worldwide practice of generative art and creative coding, as well as the adoption of these practices in both education and across creative industries. We illustrate these findings with concrete examples of repositories and contributor profiles around the globe, spanning the spectrum of code-based art, from visual to sound art. These results lay the groundwork for educators, researchers, and artists who investigate code and creative practices to  sustain diversity in the construction of our digital world.
\end{abstract}

\section{Introduction}

Since the early days of computers and software, artists have experimented with the creative potential of these media~\cite{paul2023digital,cotton_radical_2024}.
Vera Molnar pioneered the use of software to systematically explore the formal possibilities and variations of an image composition, shape, color, or scale, as opposed to the popular belief that ``art simply happens''~\cite{molnar1975toward}.
S{\'y}kora used software to explore multiple combinations to assemble abstract geometrical shapes~\cite{sykora1970computer}, while Cornock and Edmonds foresaw the role of software for interactive artworks~\cite{cornock1973creative}.
These early adopters of code as an artistic medium appreciated the high precision and power of computers, leveraging their ability to methodically and rapidly generate variants of an image for creative purposes. 
Six decades after those initial experiments, the use of code for artistic ends has evolved from pure formal exploration into a wide range of art practices. Artists from all over the world build software systems that autonomously perform a part of artistic creation at a rate and in a space vastly different from our direct perceptual experience~\cite{mccormack2001}.

\begin{figure}[h]
     \centering
     \begin{subfigure}[b]{0.48\textwidth}
         \begin{lstlisting}[language=JavaScript, escapechar=!, linewidth=\textwidth]
let numRect = 5;
let marco = 20;
function draw() {
 translate(marco, marco);
 for (let i=0; i<numRect; i++) {
  for (let j=0; j<numRect; j++) {
   let col = Math.floor(random(2));
   switch(col) {
    case 0:
     fill(0,100,60,170); break;
    case 1:
     fill(0,100,100,170); break;}
   let move = random(7);
   let inX = move+(width-2*marco)/numRect*i;
   let inY = move+(height-2*marco)/numRect*j;
   rect(inX, inY, width/numRect-10, height/numRect-10);}}
noLoop()}
        \end{lstlisting}
     \end{subfigure}
\begin{subfigure}[b]{0.42\textwidth}
         \centering
    \includegraphics[width=\textwidth]{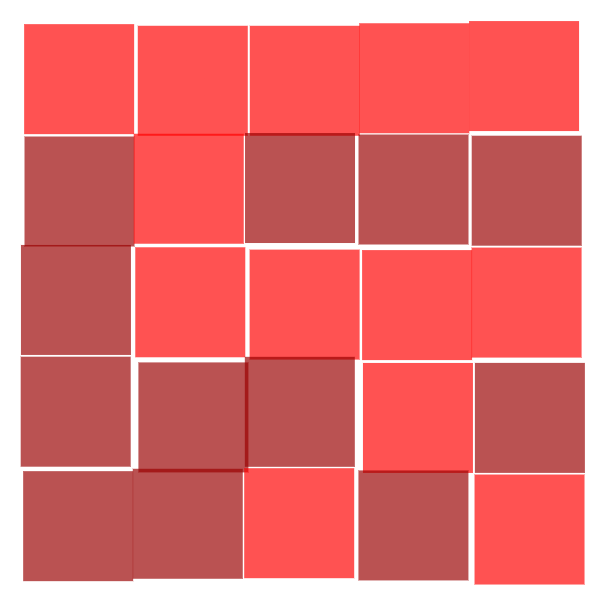}
\end{subfigure}     
\caption{Example of a program that generates art. Each time the program is executed it draws a unique artwork. It is inspired by \href{https://gazelliarthouse.com/artists/220-vera-molnar/works/8146-vera-molnar-25-carres-1991/}{25 carrés} (\href{http://www.veramolnar.com/}{Vera Molnar}, 1991). The  code was developed in 2016, it is publicly available \href{https://github.com/azurahj/electro-sounds_1516-sem2/blob/master/assignment-a/sketches/final/vera_molnar_AJ/vera_molnar1.pde}{on GitHub} and archived at \href{https://archive.softwareheritage.org/browse/content/sha1_git:176c0931b6f99d48559f7a99e97de1a74472c097}{\small swh:1:rev:2d5ba602e04d55664070b7d6f6297123021fcab5}.}
\label{fig:example-generative}
\end{figure}

Open-source software has played an essential role in the growth of these artistic practices. In electronic music communities, the main projects were Puckette's development of \texttt{Pure Data} in the early 1990s~\cite{puckette1996pure}, as well as the first versions of \supercollider in 1996~\cite{mccartney2002rethinking}. In the design and visual arts communities, Maeda's \emph{Design by Numbers}~\cite{maeda2001design} led to the development of \processing by Reas and Fry in the early 2000s~\cite{reas2007processing}. This supported the emergence of a whole new community of visual artists through an open-source language and editor tailored to their needs for the exploration of visual and interactive arts. \processing, \OF~\cite{noble2009programming}, and \pfive~\cite{mccarthy2015getting} are open-source libraries that have played a key role in the development and sustained practices of generative art~\cite{boden2009generative} over the past two decades. These libraries have also been used by software developers to create artistic, expressive, or engaging outputs. Such practices go beyond pure functionality and utility and are commonly referred to as creative coding~\cite{peppler2005creative}.

\autoref{fig:example-generative} illustrates a concrete example of using code to create a generative artwork. At the top of the figure, there is a program that draws a different image every time it runs. This program uses the \processing library. The \texttt{draw()} function contains two nested loops that divide the canvas into a grid, and \texttt{move} is a random variable that defines an offset to draw each square.

Our work analyzes the diversity of practices in public code repositories that host such pieces of generative art and creative coding projects. In order to make our search for art-related code as inclusive as possible and to avoid excluding code that is no longer available online, we mine code repositories from the \swh archive \cite{ipres-2017-software-heritage} instead of selecting a specific hosting platform such as GitLab or GitHub. Furthermore, to distinguish art-related code repositories from others, we focus our study on repositories that explicitly use an open source library which is dedicated to generative art.

We collect \nborigins  art-related code repositories from \swh. This remarkable number of repositories is evidence of the vibrant international activity surrounding generative art and creative coding. It is also a challenge for analysis. Accordingly, we define four research questions that explore this dataset on four different scales. First, we analyze the whole dataset to determine which code hosting platforms art contributors use, and we inspect which repositories are still online. Second, we select a subset of 10\% of the repositories to study the geographical distribution of their main contributors. Third, we collect 1\% of the \pfive repositories and use an LLM to automatically characterize the creative practices found in \pfive sketches. Finally, we focus on the top 42 accounts that hold the largest number of repositories, and manually analyze their content and purpose with respect to generative art and creative coding.

Previous work at the intersection of software engineering and art has either proposed novel techniques to support generative art practices~\cite{islam2025trigraph, fredericks2023generative, angert2023spellburst} or relied on interviews and case studies to empirically examine the key features of software engineering systems for the arts~\cite{ossta18, trifonova2008software}. In contrast and as a complement to this work, we present the first large-scale empirical study of open-source repositories for the arts. Our key results are as follows: \pfive and \processing play an essential role in generative art and creative coding; \swh has already saved \nbdead repositories that disappeared after hosting services shut down or authors removed them; the analysis of \nbrqtwosample repositories allowed us to unveil contributors in  \nbuniquecountries countries; the automatic analysis of \nbpfiverqthree \pfive sketches reveals that creative coders take advantage of the library's versatility to develop very diverse works, from randomized still images to interactive audiovisual pieces; and the systematic analysis of the top 42 contributing accounts confirms three main purposes for creative coding in public repositories: computer science education, art education, and art. These new observations provide evidence and case studies for researchers in human aspects of software engineering, the conservation of creative code and software engineering for the arts, as well as for practitioners in the areas of art education, computer science education, and open-source development for the arts.
The key contributions of this work are as follows:
\begin{itemize}
    \item We present the first large-scale empirical study of \nborigins public code repositories related to generative art and creative coding.
    \item Evidence that (i) a portion of this code has already disappeared and archives such as \swh play a key role in its conservation; (ii) practitioners who use code as an artistic and creative medium are found all around the world; (iii) \pfive users take advantage of the library's versatility to develop a wide variety of works; and (iv) art-related code can have three main purposes: education in art, education in computer science, or the development of art installations and new media artworks;
    \item An open dataset of source code repositories related to generative art and creative coding on \href{https://zenodo.org/records/20184094}{Zenodo} with all code and data for this study on \href{https://github.com/sparkrew/art-in-humanitys-code}{GitHub}.
\end{itemize}

\section{Related Works}

\textbf{Empirical studies of Computational Art Practices}
Several studies have examined how artists use algorithms, programming environments, and computational systems as creative media. Wu et al.~\cite{wu2024survey} review artificial life algorithms that generate visual art and propose a taxonomy for analyzing such artworks; while Verano et al.~\cite{verano2023art}  interview five creative coders in order to understand their experiences across different stages of the creative development process. Other studies focus more on code practices, as is the case for Subbaraman et al.~\cite{subbaraman2023forking}, who analyze how creative coding artists reuse and transform existing code in OpenProcessing; and in our previous work ~\cite{baudry25}, we construct a dataset of open-source projects used by media artists. Related work has also explored software practices in generative music. Islam et al.~\cite{IslamEH24} analyze the Pure Data (PD) visual language to identify programming patterns that can support users' development practices, while Bogdan et al.~\cite{BogdanMM25} mine GitHub repositories  of open-source Virtual Studio Technology (VST) plugins, to extract the engineering characteristics of these projects. Payne et al.~\cite{payne2025exploring} study the importance of online forums for live coders, focusing on how artists share code, discuss challenges, and support each other in Tidal Cycles and Sonic Pi communities.
These studies highlight the importance of examining how artists use emerging technologies to create new forms of digital arts. We contribute to this body knowledge with a study at a very large scale and a focus on practitioners and their intention when using code.

\textbf{Empirical Studies of Open-Source Contributors}
Several works focus on open-source contributors.
The growth and availability of public code repositories have facilitated empirical studies of open-source contributors, their roles, practices, and community dynamics. Milewicz et al.~\cite{milewicz2019characterizing} combine repository-level metrics with interviews to describe the roles that contributors play in open-source scientific software projects. 
Trinkenreich et al.~\cite{trinkenreich2024unraveling} survey developers and contributors to examine the circumstances that shape members' sense of belonging in a team, identifying work appreciation and psychological safety as fundamental factors. Other studies focus on diversity within open-source communities. Hyrynsalmi et al.~\cite{hyrynsalmi2025making} emphasize the need to make software communities more diverse and inclusive, especially for underrepresented groups, while Vasilescu et al.~\cite{vasilescu2015gender} examine the positive correlation between gender diversity and team productivity. Our work contributes to this body of knowledge with a special focus on art-related code repositories, and with novel results about the geographical distribution of creative open-source contributors.

\textbf{Software tools for artists and creative programmers}
Tools for artists and programmers are still being created in order to enhance their experience with code for creative purposes. Angert et al.~\cite{angert2023spellburst} introduce Spellburst, an LLM-powered creative coding environment to generate and explore variations of visual programs through prompt-based interaction, branching, and merging. 
Fredericks et al.~\cite{fredericks2024generativegi} investigates how evolutionary algorithms can amplify the generation of artwork variants.
Li et al.~\cite{li2020supporting} propose Demystified Dynamic Brushes (DDB) to facilitate the exploration of how numerical inputs influence the state of visual artworks, and Sacadura \cite{sacadura2026} propose a solution to explore the space of parameters combinations. 
Islam et al.~\cite{islam2025trigraph} propose TriGraph, a graph-based probabilistic model for code completion in Pure Data. 
Our observations about the purposes and practices of generative art and creative coding can inform future efforts to expand the use of code for artistic creation.

\textbf{Software engineering challenges in Software-Based Art}
At the intersection of software research and generative art practices, several studies have examined the software engineering challenges that artists face while creating their pieces~\cite{li2021we}. Trifonova et al.~\cite{trifonova2008software} document the specific engineering challenges regarding requirements, architecture and design, testing, validation, and maintenance that occur when developing new media artworks.
These challenges are also discussed in community events such as the Open Source Software Toolkits for the Arts (OSSTA): a Convening (2018)~\cite{ossta18} and the Open Source Arts Contributor’s Conference (OSACC, 2023)~\cite{ossart23}. In these events, the attendants articulate and address the challenges of sustainability, maintenance, accessibility and funding for open source software for and by artists \cite{coleman2026}.

\textbf{Creative coding and education}
Sandberg et al.~\cite{sandberg2019creative} show that educators have adopted creative coding as a way to teach programming, while at the same time developing creativity and innovation. Woo et al.~\cite{woo2022problem} examine how students use computational thinking and computer science concepts when coding animated narratives. 
Greenberg et al.~\cite{greenberg2012creative} use \processing to teach programming concepts through creative activities.

The same environments are used to teach programming to art students. Levin and Brain \cite{levin2021code} have collected a series of possible assignments, together with interviews with educators who share their experiences teaching programming to art students. 
Sykes~\cite{sykes2023seeing} examines how creative coding is taught in art schools, and proposes new pedagogical approaches to  help students better grasp programming concepts for their artistic practice.  
McNutt et al.~\cite{mcnutt2025slowness} analyze how tools and technologies for making art with code affect students’ learning and artistic development. Knochel~\cite{knochel2015if} argues that computational thinking and critical digital making should be integrated into art curricula, where code is an essential pillar of the creative expression. Similarly, Dufva et al.~\cite{dufva2018art} explore how to bridge the gap between artistic expression and computational education in order to have more engaging classes for students.

\section{Building A Corpus of Art-Related Public Code Repositories}
\label{sec:corpus}

We mine source code repositories related to arts from the \swh archive~\cite{ipres-2017-software-heritage}. In this section, we introduce our method and curation process for selecting the repositories for our study.

\subsection{Signals of art in source code repositories}

The key challenge to mine art-related code is to  distinguish between source code that has a creative intention from source code that has any other motivation. Generative art and creative coding can be developed in any programming language and totally independently of specific libraries or art-related toolkits. Consequently, any C or Rust project in \swh could be related to art or have a totally different purpose. We address this challenge with a conservative strategy: we mine projects that explicitly use one of the libraries that are meant for generative artists.
Such libraries can focus on visual arts (e.g., \href{https://processing.org/}{Processing}, \href{https://openframeworks.cc/}{openFrameworks}, or \href{https://nannou.cc/}{nannou}), or on sound synthesis (e.g., \href{https://supercollider.github.io/}{SuperCollider} or \href{https://faust.grame.fr/}{Faust}). Some software libraries focus on live performances (e.g. \href{https://tidalcycles.org/}{Tidal} or \href{https://hydra.ojack.xyz/}{Hydra}). Other libraries provide artists with graphical programming environments (e.g. \href{https://derivative.ca/UserGuide/TouchDesigner}{TouchDesigner}, \href{https://puredata.info/}{Pure Data} or \href{https://cycling74.com/products/max}{Max}). Some artists also develop on top of 3D or game engines such as \href{https://www.blender.org/}{Blender}, \href{https://unity.com/}{Unity} or \href{https://www.unrealengine.com/}{Unreal}. 
Yet, to support the mining process at the scale of the hundreds of millions of repositories in \swh, occupying more than 2\,PiB of raw space, we determine the usage of one of these libraries only by the presence of specific files or file extensions. It is not feasible to check file \emph{content} at this scale.

Based on these constraints, we consider a code repository to be art-related if it matches one of the ``signals'' documented in \autoref{tab:ecosystems}.
Our mining process focuses on seven software libraries and programming environments, which artists use to create and produce software-based artworks. \pfive and \OF, shown in the top two rows of \autoref{tab:ecosystems}, are popular JavaScript and C++ libraries, respectively, mainly used for visual arts. For these two ecosystems, the presence of specific files in a source code repository indicates their use. To use \pfive, artists and creative coders can store either the \pfive bundle or its minified version directly in the repository. In the case of \OF, the files named \texttt{ofApp.cpp} and \texttt{ofApp.h} are conventionally used to structure/build an \OF project.\footnote{see \href{https://openframeworks.cc/ofBook/chapters/how_of_works.html}{How openFrameworks works}} 

We detect the usage of the 5 other libraries and environments based on file extensions. \processing is a very popular Java library for artists. While a regular third-party dependency can be used with Maven or Gradle, artists and creative coders can also develop code on top of \processing using a dedicated code editor, in which case the files for the artwork are saved with the extension ``.pde''. \texttt{\supercollider} is an audio programming language built on top of C++ and all \texttt{\supercollider} files have the ``.scd'' extension. The last 3 environments are node-based programming environments. \texttt{TouchDesigner} is a proprietary runtime, but the networks, configurations, and operators for audiovisual installations are saved in files with the ``.toe'' and ``.tox'' extensions and can be shared as open source. \texttt{nodebox} and \texttt{vvvv} are node-based environments for graphical and live programming, saved in specific ``.ndb'' and ``.v4p'' files.

\begin{table}
\small
\captionsetup{font=small}
\caption{Programming frameworks for the arts and their corresponding filename patterns}
\label{tab:ecosystems}
\centering

{\scriptsize
\tabcolsep1.7pt
\def\arraystretch{1.2}
\begin{tabular}{llr}
\toprule
Art Programming Env. & Language & Filename pattern\\
\hline
\href{https://github.com/processing/p5.js}{\pfive}&JavaScript&``p5.js'', ``p5.min.js''\\
\href{https://github.com/openframeworks/openFrameworks}{\OF} &C++&``ofApp.cpp'',``ofApp.h''\\
\href{https://github.com/processing/processing4/}{\processing} &Java&``.pde''\\
\href{https://github.com/supercollider/supercollider}{\texttt{\supercollider}}&C++&``.scd''\\
\href{https://derivative.ca/UserGuide/TouchDesigner}{\texttt{TouchDesigner}}&node-based + Python&``.toe'', ``.tox''\\
\href{https://github.com/nodebox/nodebox}{\texttt{nodebox}}&node-based + Python or Clojure &``.ndbx''\\
\href{https://github.com/vvvv}{\texttt{vvvv}}&node-based + C\#&``.v4p''\\

\bottomrule
\end{tabular}
}
\end{table}

\subsection{Finding Art Code Repositories in \swh}

\swh (SWH)~\cite{ipres-2017-software-heritage} archives software source code from public sources, including major collaborative development platforms such as GitHub and multiple GitLab instances.
The source code in \swh is archived with its full development history, captured by modern distributed version control systems (VCS) such as Git; if an archived Git repository was to disappear, it could be recreated from the archive.
The \swh data model is a single, fully de-duplicated Merkle graph structure, which corresponds intuitively to the global VCS of public code.
At the time of writing (May 2026), \swh has archived more than 400 million code repositories from more than 5,000 code hosting platforms, totaling almost 30 billion unique source code files and occupying more than 2\,PiB of storage.

This is our main motivation to search for art-related code in \swh instead of GitHub or GitLab.com only. First, it includes art-related code repositories that are currently unavailable online, e.g., because the hosting service shut down, but are archived on \swh. Second, it gives access to source code that was shared before the launch of GitHub (2008), e.g., \processing art works published in the early 2000s on platforms such as \texttt{SourceForge} (launched in 1999). Indeed, the archive contains code that predates its own creation. Such code is obtained either via successive migrations through online repositories, or dedicated digital archaeology action. For example, the archive includes the code of the Apollo space missions Guidance Computer (AGC), a milestone of our software heritage developed in the late 1960s under the leadership of Margaret Hamilton.

We search for art-related repositories within this large volume of code via the filename patterns listed in \autoref{tab:ecosystems}.
We start from the \swh graph dataset~\cite{pietri-2020-swh-graph-dataset}, version 2025-05-18,\footnote{Retrieved from \url{https://datasets.softwareheritage.org/graphs/compressed/\#2025-05-18-compressed}} using its compressed representation~\cite{boldi-2020-swh-graph-compression}.
This allows us to load into memory (on a beefy server) the entire VCS history of all repositories archived by \swh and to mine them via a Rust API based on WebGraph~\cite{fontana-2024-webgraph-rs}.

In order to identify repositories that contain art-related filenames, we: (1) iterate over all repositories, called ``origins'' in \swh terminology; (2) inspect the most recent archived snapshot of each repository; (3) consider only the most recent commit on the main branch of the repository, the equivalent of Git \texttt{HEAD} at archival time; and (4) recursively list the files and directories of that commit starting from the root directory, with a recursion limit of 1,024 directories to avoid Git/tar bombs.

If, at the end of the recursion, we encounter one or more paths matching the patterns of \autoref{tab:ecosystems}, we flag the repository as potentially containing art-related code and record it along with the mined metadata, including the repository URL, the number of matches, and, for each match, the path name and intrinsic identifier.
Each matching repository is saved as an NDJSON entry, as shown below. The example is edited and reformatted for brevity and readability:

{\scriptsize
\begin{verbatim}
{
  "ori_url": "https://github.com/imclab/oauthp5",
  "match_count": 13,
  "matches_info": [
    ["swh:1:cnt:a92975aa3a6d9bd0d37d7de0606e8a1b2da76c39",
     "OpenPathsExample.pde"],
    ["swh:1:cnt:32b2e44f256e7b278516e7f30792d45bdc4b5abe",
     "Twitter3LegExample.pde"],
    ...
  ]
}
\end{verbatim}
}

We release the implementation of this mining step as part of the open-source \swh repository-mining toolkit.\footnote{\url{https://gitlab.com/zacchiro/swh-repo-mining/}, binary \texttt{find-files}. We use commit \href{http://archive.softwareheritage.org/swh:1:rev:d9412b0d5c9ebe4c907b27aa84084382309ce92e}{swh:1:rev:d9412b0d5c9ebe4c907b27aa84084382309ce92e} in our experiments.}

At the end of this mining step, we retrieve \nboriginsinitial repositories that match one or more art-related code signals in their most recently archived revision in the \swh archive.

\subsection{Consolidated Dataset of Source Code Origins}

\begin{figure}
    \centering
    \includegraphics[width=0.9\linewidth]{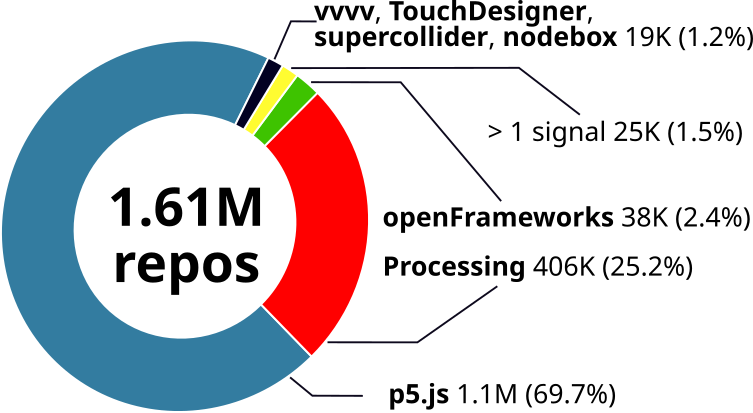}
    \caption{Distribution of the art software libraries among the \nborigins source code repositories mined in \swh.
    }
    \label{fig:signal_distribution_donut_chart}
\end{figure}

From the raw output of the  mining process described above, we perform the following steps to clean the dataset. 
First, we normalize and de-duplicate repository URLs to ignore differences in protocol (e.g., \texttt{http} vs.\ \texttt{https}) and optional \texttt{.git} suffixes, and case sensitivity. 
Second, we noticed an ambiguity in our file extension signals:  \texttt{.scd} files may correspond to either the \supercollider source code or Unix man pages. To triage these cases, we implement a custom script using two sets of keywords that are certainly present in \supercollider and in Unix \texttt{man} pages. For example, files containing terms such as \texttt{NAME}, \texttt{SYNOPSIS}, or \texttt{DESCRIPTION}, among others, are classified as \texttt{man} pages, whereas files containing terms such as \texttt{SynthDef}, \texttt{Synth}, or \texttt{ControlSpec}, among others, are classified as \supercollider source code files.

The resulting dataset, which forms the foundation for our empirical study, includes \nborigins repositories. \autoref{fig:signal_distribution_donut_chart}  shows the distribution of signals in these repositories. The majority of the repositories match the \pfive signals (69.7\%), followed by \processing (25.2\%) and \OF (2.4\%). The remaining repositories match more than one signal (1.5\%) or match \texttt{\supercollider}, \texttt{TouchDesigner}, \texttt{nodebox}, or \texttt{vvvv} signals (1.4\%).

\section{Methodology}
\label{sec:methodology}

\newcommand{\rqprov}{Where is art-related code hosted and how much of it disappeared over time?}

\newcommand{\rqartists}{What is the geographic distribution of contributors to art-related code?}

\newcommand{\rqcharacteristics}{Which characteristics stand out in artworks made with \pfive?}

\newcommand{\rqtop}{For what purpose do the top \nbtopusers contributors in our dataset develop or use art-related public code?}

\subsection{Research Questions}

\begin{figure}
    \centering
    \includegraphics[width=0.9\linewidth]{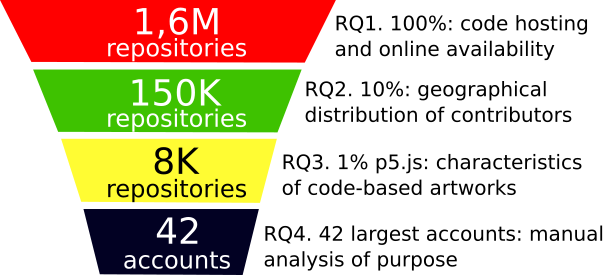}
    \caption{Four research questions exploring the dataset at different scales}
    \label{fig:funnel-methodo}
\end{figure}

We study the \nborigins repositories at different scales, as illustrated in \autoref{fig:funnel-methodo}, and answer the following research questions.

\textbf{RQ1: \rqprov}

For this first research question, we analyze the complete dataset. We determine where the artists and creative coders host their public source code. We also investigate to what extent all the repositories that we have mined in the \swh archive are still accessible online. Indeed, some code hosting platforms have disappeared over the years, and they might have hosted art-related code. It is also possible that contributors decide to delete some work. In both cases, our analysis reveals the role that an archive like \swh plays in preserving source code for the arts and creative coding.

\textbf{RQ2: \rqartists}

Open-source software is widely accessible, and anyone in the world can access libraries to practice generative art and creative coding. With this research question, we investigate to what extent these practices are indeed found all around the world.

\textbf{RQ3: \rqcharacteristics}

Open-source libraries for generative art and creative coding are versatile and can support a wide variety of artistic expressions. For example, it is possible to generate static, abstract graphical compositions, as well as to control interactive audiovisual installations.
With RQ3, we aim to unveil these various practices among the users of the most popular generative art library, \pfive.

\textbf{RQ4: \rqtop}

There are many reasons why people contribute source code with creative coding libraries. Understanding these different purposes provides  qualitative insights about the motivations of users and can inform further development of these practices. This research question focuses on the \nbtopusers most prolific accounts in our dataset and the analysis of their main purpose.

\subsection{Methodology for RQ1}

For this question, we collect two pieces of metadata for the complete set of \nborigins repositories.
First, we determine on which platform each source code repository is hosted. We extract this information from the URL of the source code repository, e.g., \texttt{gitlab.com} or \texttt{git.sr.ht}. Second, we determine if the repository is still available online. Since we have collected our dataset from the \swh code archive, it is likely that some repositories still present in the archive are no longer accessible online from their original hosting place, e.g., if the authors deleted it or if the hosting service shut down. 
The status of each URL is determined by sending HTTP requests (to the original hosting place) and interpreting the response codes. Status codes below 400 are considered online, whereas 4xx or 5xx indicate that the resource is no longer available. 
The first metric is an indicator of where code is distributed from, while the second one reveals challenges related to the conservation of this artistic heritage. 

\subsection{Methodology for RQ2}

For this question, we randomly sample 10\% of the online GitHub repositories in our dataset.
We focus on GitHub as it hosts the most repositories and its API allows us to obtain user location information in a consistent way.
To capture as many different contributors as possible, we make sure that each repository in the sample comes from a different GitHub account.
In addition, we check that the selected repository is still online.

Following this process, we build a sample of \nbrqtwosample GitHub repositories. Key statistics about this sample are given in \autoref{tab:contrib-locations}.
We fetch the main contributor (i.e., the one who authored the most commits) for each repository, leading to  \nbuniquecontributors unique contributor profiles.
Then, we collect the location declared on their GitHub profile.
On GitHub, the location field is optional; in our sample, \nbnonemptylocations contributors provide a non-empty location string.
Since location values are free form with no validation, these location entries are not standardized, e.g., ``Montréal, QC'', ``Montreal, Canada'', and ``MTL'' are valid, different strings that designate the same location.
People may describe their location in different ways, such as a neighborhood, city, country, or even something unrelated to a real place (e.g., ``Tatooine'', ``my basement'', or ``the Internet'').
This creates inconsistencies and possible errors for mapping the data. 

To normalize location values and obtain actual geo-locations, we use the Geoapify Location Platform\footnote{\url{https://www.geoapify.com}}.
For each contributor with a non-empty location field, we invoke the Geoapify API to convert the text into a consistent location format.
When the input can be successfully matched to a real-world location, the API returns a valid, standardized location with corresponding GPS coordinates.
After this step, we obtain  \nbreposwithvalidlocation valid locations out of the \nbrqtwosample entries (9\%).
When a location is resolved to a place other than a country, such as a city or a state, we invoke the Geoapify API again with the location's GPS coordinates to determine the corresponding country. With the Bertin.js library\footnote{\url{https://github.com/riatelab/bertin}}, we visualize the different locations on a world map using a polyhedral projection~\cite{brehm2017polyhedral}.

\begin{table}
\small
\captionsetup{font=small}
\caption{Consolidating locations of main contributors}
\label{tab:contrib-locations}
\centering

{
\tabcolsep1.7pt
\begin{tabular}{l@{\hskip 0.5in}r}
\toprule
Original number of repositories in the dataset  & \nbrqtwosample \\
Number of unique main contributors & \nbuniquecontributors\\
Number of contributors that declare a location & \nbnonemptylocations\\
Number of declared locations resolved & \nbreposwithvalidlocation\\
\bottomrule
\end{tabular}
}
\end{table}

\subsection{Methodology for RQ3}

\begin{table}
\small
\captionsetup{font=small}
\caption{Curating a dataset of single-file \pfive sketchers}
\label{tab:pfive-sketches}
\centering

{
\tabcolsep1.7pt
\begin{tabular}{l@{\hskip 0.5in}r}
\toprule
Original number of repositories in the dataset  & \nbrqthreesample \\
Number of \pfive sketch files & \nbrawpfivesketches\\
Number of \pfive sketch files under study & \nbpfiverqthree\\
\bottomrule
\end{tabular}
}
\end{table}

First, we select a random 1\% sample of repositories that use \pfive and are hosted on GitHub. Then, we search the repositories for source code files that implement a piece of art or creative coding. We refer to these files as \emph{sketch files}. In \pfive, a sketch file is either an embedded JavaScript script in an HTML file, or a standalone JavaScript file.
We check whether any HTML file in the repository includes the \pfive special \texttt{setup()} function. In such cases, we consider the HTML file to be the sketch file.
If no such HTML file is found, we look for folders containing an HTML file together with one or more JavaScript files. Then, we check the name of each JavaScript file to determine whether it corresponds to one of the numerous \pfive additional libraries. For this, we use a predefined list of known \pfive libraries per the online documentation.\footnote{\url{https://p5js.org/libraries/}}  If a JavaScript file is not identified as a library file, we then check whether it contains a \texttt{setup()} function. If it does, we consider that JavaScript file to be the sketch file. In order to focus on sketches that actually perform some creative processes, we ignore the sketches that invoke fewer than three \pfive functions, resulting in \nbpfiverqthree sketches for analysis.
Key metrics regarding the sample size and the collected sketch files are presented in \autoref{tab:pfive-sketches}.

Second, we define the characteristics of sketches that we wish to capture, in order to classify \pfive usage practices. For this, we are inspired by the framework for understanding generative art introduced by Dorin and colleagues~\cite{dorin2012}. 
The \matproc labels capture elements of the sketch related to the entities and actions it manipulates in order to perform the artwork: \procaudio, \procimage, \proctext  capture the fact that a sketch can process preexisting material to display it as part of the work or to extract data from it to include in the generative process; \synaudio, \synimage, \syntext are values related to the fact that a sketch can use code to generate and evolve totally new material; \randomness captures the fact that the sketch relies on randomness  to increase the level of autonomy given to the machine in the generative process. The \interaction label can be \interactive or \noninteractive and captures whether the sketch has code related to user interactions. 
The \outcome labels capture the intention of the code regarding the audience's perception of the work.
The outcome can be \visual if the code displays images or videos, generated or processed from existing sources; \auditory outcome is when the work streams sound, generated or replayed from previous sources; and the outcome is either \static or \timebased. The label ``time-based'' is borrowed from the art conservators' terminology to refer to a work that unfolds over time \cite{frohnert2017time}.

Third, we use an LLM to automatically characterize our \nbpfiverqthree sketches, according to the classification of \autoref{tab:framework}. 
In order to design a prompt for this labeling task, we first manually label 74 sketches. Two authors label them, compare their labels, and ask a third author to adjust in case of disagreement. Then, we design a first prompt and query the LLM to label the same 74 sketches. 
The first version of the prompt includes simple descriptions for each label created by the authors. It contains a one-sentence description for each characteristic, \matproc, \interaction and \outcome, as well as a fixed JSON schema for the expected output format for further processing.\footnote{The initial version of the prompt is available as part of our reproduction package: \url{https://github.com/sparkrew/art-in-humanitys-code/blob/main/rq3/scripts/initial_prompts.py}
An iterative process of prompting the model to generate labels, analyzing the true and predicted label distribution, as well as of comparing the labels inferred by the model with their respective p5.js source code reveal the need to strengthen the descriptions of each characteristic in order to meet expectations. 
Therefore, we substitute the initial descriptions with much stricter and comprehensive definitions, using concrete p5.js function examples when applicable, explicit decision rules, and edge-case checks in order to create the final version of the prompt.}
We validate this refined prompt with another sample of 50 manually labeled sketches. 
This last round led to minor adjustments of the prompt \footnote{The final version of the prompt is available here: \url{https://github.com/sparkrew/art-in-humanitys-code/blob/main/rq3/scripts/prompt.py}}. 
The prompt is organized as follows. First, it defines the expected output format as a JSON object with three fields: \matproc, \interaction, and \outcome. Second, it specifies the allowed labels for each characteristic and explains how each label should be assigned, including definitions and examples. Finally, it includes decision rules for ambiguous cases, such as the detection of \randomness and the classification of \timebased behavior. 

To analyze the complete set of \nbpfiverqthree sketches, we pass the prompt and the source code of a \pfive sketch to the model. The model output is a list of labels for each characteristic of the sketch. In case of failure or ill-formed output, we run the same prompt and file again one more time.
We use the Qwen3-Coder-30B-A3B-Instruct-FP8 \cite{qwen3technicalreport} model for inference. The model was deployed on an NVIDIA H100 GPU with 80\,GiB of VRAM, and 8\,GiB system RAM, using the Rorqual compute cluster, operated by Calcul Québec,\footnote{\url{https://www.calculquebec.ca/}} a regional partner of the Digital Research Alliance of Canada.\footnote{\url{https://www.alliancecan.ca/}}

\begin{table}
\small
\captionsetup{font=small}
\caption{Three key characteristics and possible values for each characteristic to classify \pfive sketches}
\label{tab:framework}
\centering

{
\setlength\tabcolsep{5.4pt}
\begin{tabular}{p{0.22\linewidth} | p{0.7\linewidth}}
\toprule
\textbf{Characteristics} & \textbf{Values} \\
\hline
\matproc &  \procaudio;
\procimage;   \proctext;  \synaudio;   \synimage;  \syntext;  \randomness \\\hline
\interaction & \interactive  or \noninteractive \\\hline
\outcome & \visual; \auditory; \newline    \static or \timebased \\
\bottomrule
\end{tabular}
}
\end{table}

\subsection{Methodology for RQ4}

For this question, we manually analyze the \nbtopusers GitHub accounts with the largest number of repositories in our dataset. We focus on these accounts, as their large number of repositories indicates a serious, sustained interest in code for art.

For each account, we collect a set of descriptors, such as the number of repositories, the type of account (individual or organization), the main programming language(s), and the activity status of the account. 
We classify an account as ``active'' if it shows activity on GitHub within the last two years, and ``inactive'' otherwise.

We also manually determine the purpose of each repository. Two authors independently conducted inductive coding of the repositories associated with these \nbtopusers GitHub accounts. Rather than applying predefined categories, they developed codes that emerge directly from the data (repository names, descriptions, code comments, README files, and GitHub profile information). This coding process aims to capture the primary purpose of the repositories and to characterize the level of activity of these accounts. After the initial coding phase, the two researchers compared their findings. In cases of discrepancy, they consulted a third author. This process resulted in three main purpose categories: \emph{art}, \emph{art education}, and \emph{computer science education}.
Together, these various metrics and assessments provide original, contextual insights about the various purposes for using and sharing code for art.

\section{Results}

\subsection*{RQ1. \rqprov}

\begin{figure}[t]
    \centering
    \includegraphics[width=1\linewidth]{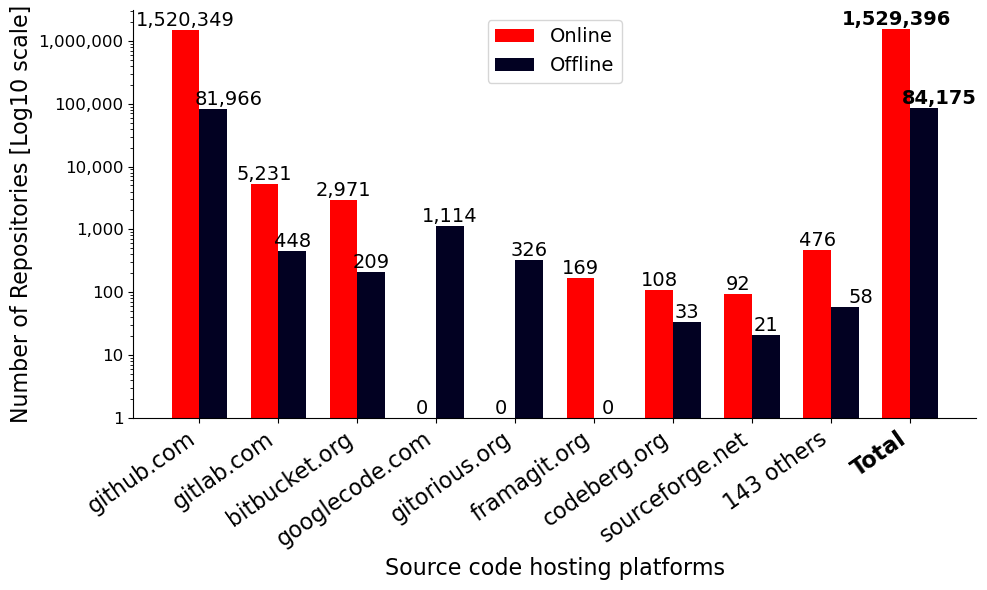}
    \caption{Distribution of code repositories on different hosting platforms. For each hosting platform, we represent the number of repositories that are accessible as of January 06, 2026 (online), as well as the number of repositories that are not accessible anymore (offline). The 143 platforms that host less than 100 repositories are merged together in the ``others'' category.}
    \label{fig:art_hosts}
\end{figure}

\autoref{fig:art_hosts} presents the distribution of public repositories for generative art, across different code hosting platforms.
Given the large discrepancies among the number of repositories, we use a $\log_{10}$ scale on the Y axis.
In the figure, we also distinguish between repositories that are still available online and those that are no longer available from their original hosting place, but remain accessible through the \swh archive. The last pair of bars, on the right of the figure, presents the total numbers of online and offline repositories.

GitHub is by far the most popular hosting platform, with $99.3\%$ of all repositories: 1.5 million online repositories and 81k offline ones.
GitLab and Bitbucket follow distantly, with 5,231 and 2,971 online, and 448 and 209 archived repositories, respectively. 

Two hosting platforms, \texttt{googlecode.com} and \texttt{gitorious.org}, only have offline repositories.  Gitorious was discontinued in 2015 when it was acquired by GitLab, which offered an opportunity to migrate the code repositories. We found 45 out of 326 URLs that matched organization and repository name in GitLab, suggesting that they were migrated from Gitorious to GitLab(.com). For example, Rhea Myers, a pioneer generative and NFT artist, migrated her project \href{https://gitlab.com/robmyers/like_that.git}{like\_that}, to GitLab. Yet, the repository has 0 commits on GitLab, evidence that it was migrated from Gitorious for preservation purposes. 
Google Code was discontinued in 2015. Out of the 1,114 repositories on Google Code, almost all matched the \processing signal, whereas only two repositories included \supercollider and one included \texttt{vvvv}.
Thanks to our mining of \swh, we could still access all of these art-related code repositories, which are now more than 10 years old.

\begin{figure*}[t]
     \centering
  \includegraphics[width=0.9\linewidth]{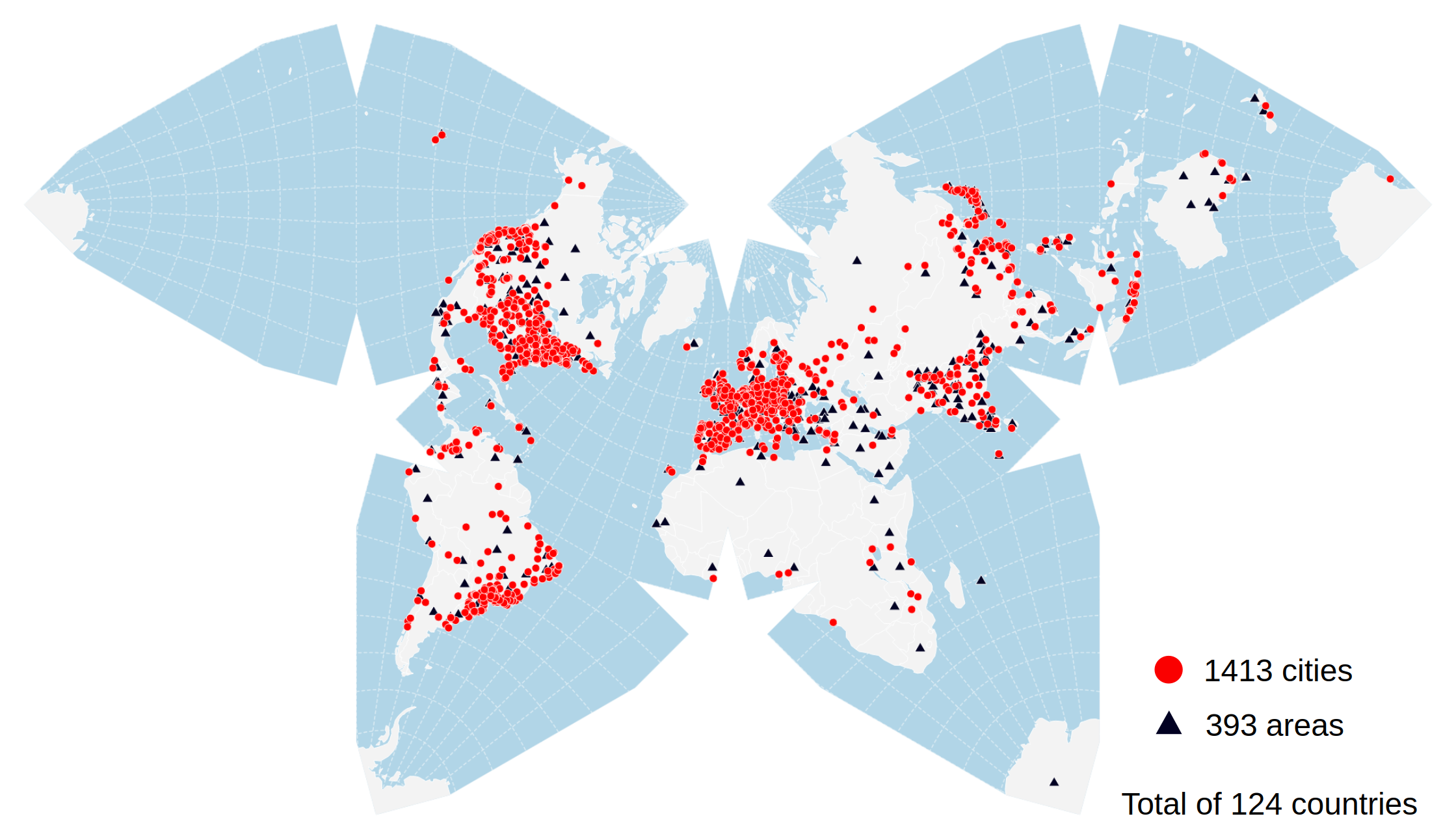}
     \caption{Map of the 1,806 geographical locations gathered from the \nbreposwithvalidlocation contributors to generative art and creative coding, who have a valid location as part of their GitHub profile. We consider valid names of cities, or names of areas, such as a county, state, or country. The 1,806 locations are in 124 different countries across all continents.}
     \label{fig:art_locations}
\end{figure*}

Codeberg is an emerging collaborative software development environment created in 2018. It has grown to host nearly 485,000 repositories as of April 2026, of which 141 are identified in our dataset as generative art projects. An example is \href{https://codeberg.org/pfzzz/MicroDexed}{pfzzz/MicroDexed}, a synthesizer firmware project that combines code and sound design. This repository created in 2018 was included in our dataset because it contains \processing files. Codeberg is often mentioned in forums as an alternative to the monoculture of GitHub. The emergence of 141 art repositories in under a decade in Codeberg illustrates this vitality of open-source creative coding beyond GitHub.

Beyond these major platforms, the remaining repositories are scattered across 143 unique hosts, with 476 online and 58 offline repositories. Out of these 143 hosts, 29 have only offline repositories. These hosts are run by public organizations, schools, makers associations or a few individuals. Most of these hosts run a local GitLab, e.g., Stanford's GitLab instance at \href{https://code.stanford.edu}{code.stanford.edu}, which hosts a number of repositories that use \processing for designing and documenting art installations.
We also found niche self-hosted forges such as the \href{https://about.gitea.com/}{gitea} forge instance of the French robotics association \href{https://git.poivron-robotique.fr/}{Poivron robotique}, or a
\href{https://forgejo.org/}{forgejo} instance for the code of \href{https://molzy.com/git/explore/repos}{Simon Molzy}.
Meanwhile, the \href{https://github.com/rekkabell}{Rek} and \href{https://github.com/neauoire}{Devine} artists duo hosts their code on  \href{https://git.sr.ht/~rabbits}{sourcehut}.

Across all platforms, 84,175 repositories, roughly 5.5\% of the total, are no longer accessible online. This highlights the importance of code preservation (not only for software-based art) and validates the choice of mining Software Heritage, rather than a traditional online development platform like GitHub, for the goals of this study.

\begin{tcolorbox}[boxrule=1pt,arc=.3em, left=2pt, right=2pt]
  \textbf{Answer to RQ1}: While a vast majority of art-related public repositories are hosted on GitHub, public organizations, makers, and artists also use a variety of alternative hosts, from GitLab to SourceHut. \swh has archived more than 80K art-related repositories that disappeared after deletion or the interruption of hosting services. 
\end{tcolorbox}

\subsection*{RQ2. \rqartists}

\newcommand{\nbcontributorsinbrazil}{934\xspace}
\newcommand{\nbcontributorsinindia}{1,055\xspace}
\newcommand{\nbcontributorsinjapan}{586\xspace}
\newcommand{\nbcontributorsinusa}{4,303\xspace}
\newcommand{\nbcontributorsincanada}{898\xspace}

In this RQ, we focus on \nbreposwithvalidlocation GitHub users with a geo-locatable location in their profile, which we have extracted from a sample of \nbrqtwosample repositories.
\autoref{fig:art_locations} shows the 1,806 unique locations that we found in these user profiles.
In the map, we distinguish between different geographic granularities that contributors use when documenting their location: either a city, or a wider area such as a region or a country.
We keep this distinction as we can precisely locate a city on a world map (\cityicon), whereas an area's icon is positioned using its centroid (\areaicon).
The map clearly illustrates a key result from our study: open-source contributors who use source code as a medium for art and creative practices are present across the entire world.
We found contributors across all continents, for a total of 124 distinct countries present in a 10\% sample of our global dataset.

\begin{figure}[t]
    \centering
    \includegraphics[width=0.8\linewidth]{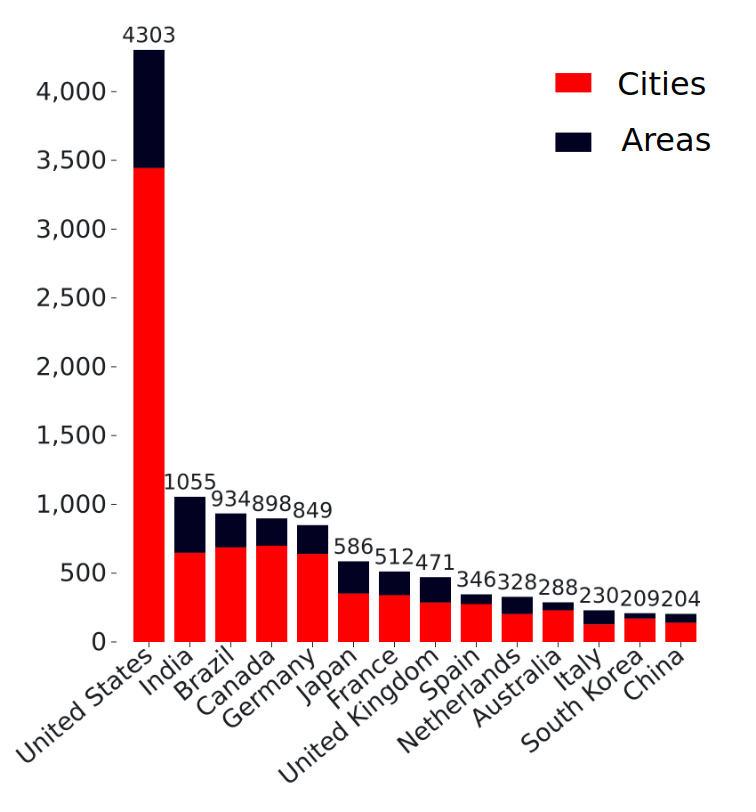}
    \caption{Countries with more than 200 georeferenced contributors to art-related public source code repositories}
    \label{fig:top_countries}
\end{figure}

\begin{figure*}[h]
     \centering
     \begin{subfigure}[b]{0.35\textwidth}
         \centering
         \includegraphics[width=\textwidth]{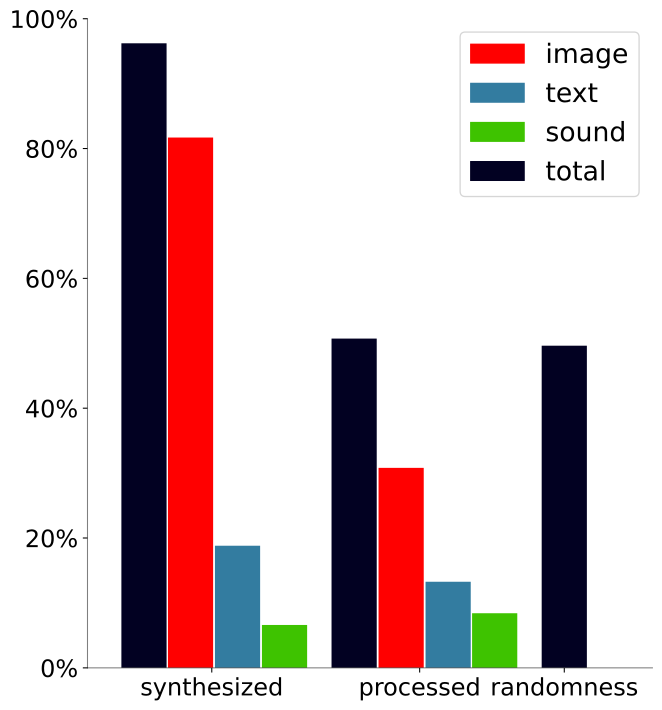}
         \caption{Material and processes}
         \label{fig:material-processes}
     \end{subfigure}
     \hfill
     \begin{subfigure}[b]{0.28\textwidth}
         \centering
         \includegraphics[width=\textwidth]{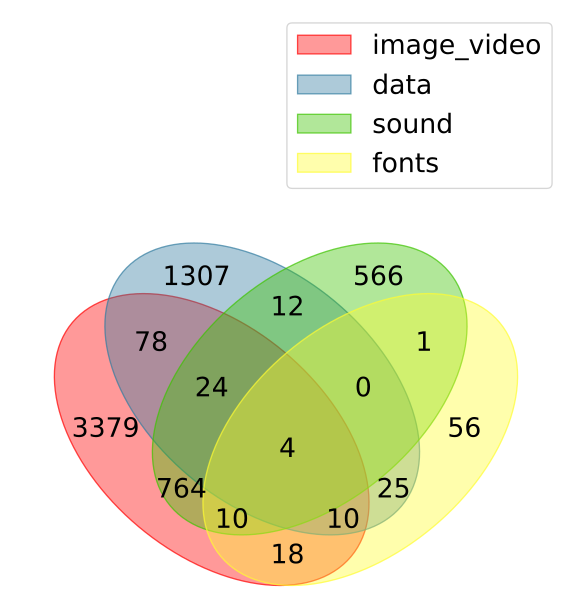}
         \caption{6,256 (34,9\%) unique sketches use external artifacts.}
         \label{fig:external-assets}
     \end{subfigure}
     \hfill
     \begin{subfigure}[b]{0.35\textwidth}
         \centering
         \includegraphics[width=\textwidth]{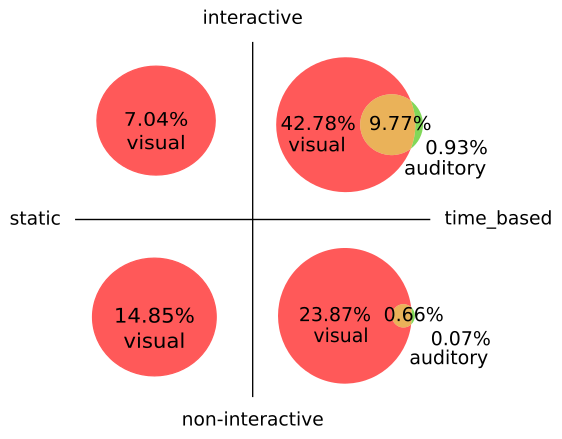}
         \caption{Interaction and sensory outcome.}
         \label{fig:interaction-outcome}
     \end{subfigure}
     
    \caption{Diversity of practices and external asset usage across a sample of \nbpfiverqthree \pfive sketches. (a) and (b) show the distribution of practices based on the LLM output, with respect to their \texttt{Material and processes} (a) and the combination between the \texttt{Sensory outcome} and \texttt{Interaction} (b). In (c) we provide a Venn diagram with an overview of sketches that use at least one external asset, grouped by image/video, data, sound or font files, to feed the piece.}
     \label{fig:llm-classification-assets}
\end{figure*}

\autoref{fig:top_countries} shows the countries for which we counted more than 200 contributors. Each bar represents the number of contributors located in a specific country, either because they declare that they are in that country or in a city or region within it. First, we observe that these contributors are present all over the world, from \nbcontributorsinbrazil contributors in Brazil to the \nbcontributorsinjapan contributors in Japan. This confirms what we previously observed on the map: there are a significant number of contributors to generative art and creative coding outside of North America and Europe. In particular, India, Brazil, and Japan rank second, third, and sixth, respectively, with \nbcontributorsinindia, \nbcontributorsinbrazil, and \nbcontributorsinjapan contributing. The United States ranks first, with \nbcontributorsinusa contributors located all across the country, from 632 in New York City and 322 in San Francisco, to 8 in Honolulu, Hawaii, and 2 in Anchorage, Alaska. In order to gain deeper insights about these contributors and their diversity, we have manually inspected two groups of contributors: the Montreal community, and the contributors in underrepresented regions.

\paragraph{The Montreal Community}

Canada has \nbcontributorsincanada contributors, despite its relatively modest population. Six co-authors of this paper have settled in Montreal in recent years and notice that 116 contributors in our dataset also position themselves in this city. Manual inspection of these 116 contributor profiles reveals that these contributors all work in various sectors of Montreal's vibrant creative industry : visual artists, light designers, sound artists, game developers, front-end developers, graphic designers, data visualization engineers, as well as students and professors in computation arts, new media arts, AI and robotics. We found a remarkable number of active profiles contributing public source code specifically for the arts. Some of them contribute within major art and creative technology organizations, such as \href{https://github.com/stephanschulz}{Stefan Schulz} working at Rafael Lozano-Hemmer's Antimodular studio, \href{https://github.com/jcelerier}{Jean-Michaël Celerier} supporting artists at la Société des Arts Technologiques, \href{https://github.com/natcl}{Nathanaël Lécaudé} building open-source solutions for audiovisual installation at Moment Factory, \href{https://github.com/nicobou}{Nicolas Bouillot} building creative technology at the cooperative Lab148. Some others develop an independent practice, such as \href{https://github.com/netherwaves}{Claire Avery}, \href{https://github.com/thomasfredericks}{Thomas Fredericks}, \href{https://davidholcer.com/}{David Holcer}, 
\href{https://github.com/mysteryDate}{Aaron Krajeski}, \href{https://github.com/pelletierauger}{Guillaume Pelletier-Auger}, 
\href{https://github.com/noiramschneider}{Marion Schneider}.

\paragraph{The Underrepresented Regions}  We identify the Caribbean, Africa, the Middle East, Central Asia, the Indian Ocean, and Iceland as underrepresented regions.
We confirm 334 contributors distributed across 20 African countries, 11 countries in the Middle East, 7 in Central Asia, 5 in the Caribbean, and 3 in the Indian Ocean, as well as Iceland.

Through our manual analysis of these contributor profiles, we encountered several active generative artists in diverse parts of the world. In Central Asia, \href{https://github.com/vonqo}{Enkh-Amar Gantulga} is an active new media artist based in Ulaanbaatar, Mongolia. His practice spans visual arts, immersive installations, and live coding. His GitHub profile is balanced between forks of creative coding projects for live coding or glitch art, and projects that he maintains, mostly for live performances. He works mostly in C++ and \OF.
In the Middle East, \href{https://github.com/zap-syr}{Aleksei Vlasov} is based in Dubai, UAE. He develops and maintains open-source projects that interact and integrate with the commercial software Disguise, a leading technology for large-scale projection mappings.
In the Indian Ocean, \href{https://github.com/vihanpereraux}{Vihan} is a new media artist based in Sri Lanka who maintains an open-source organization consisting of creative developers, artists, and researchers. On GitHub, he contributes to creative coding repositories in \pfive and generative art projects in Python. 
In Iceland, all 10 contributors in our dataset are active in sound art or research related to sound. For example, \href{https://github.com/jarmitage}{Jack Armitage} is an electronic music composer working in C++ and \supercollider, as well as in Haskell and \texttt{Tidal Cycles} for live coding. His repositories are reusable libraries for code synthesis in different contexts, e.g., embedded on a Bela board, as well as actual pieces of electronic music, such as this repository for live coding \href{https://youtu.be/sNj-I2pZwX8}{Charli XCX's Vroom Vroom}.

We also found a number of profiles related to graphic design, front-end development, interaction and Arduino development with \processing, as well as creative coding for programming education. Most of the contributors in the Caribbean and Africa are engaged in these activities. For example, \href{https://github.com/Sarah-Marion}{Sarah Marion} from Nairobi, Kenya, uses the \pfive library to design web interfaces. In Dar es Salaam, Tanzania, the high school \href{https://github.com/stemloyolatz}{STEM Loyola} teaches programming through \pfive startup challenges. \href{https://github.com/AbelWondafrash}{Abel Wondafrash} from Ethiopia is one of the many African engineers who leverage \processing to build interactive and embedded systems with Arduino boards.

\paragraph{The Case of Nonexistent Locations}
Answering this research question, we have discussed the \nbreposwithvalidlocation valid locations, which we gathered out of the \nbnonemptylocations profiles that indicate a location. Looking into the variety of nonexistent locations used by creative coders could also reveal novel insights about these practitioners. For example, the audiovisual artist \href{https://github.com/neauoire}{Devine Lu Linvega} indicates they are located in \texttt{"Nepturne 7757"}, while they develop open-source technology for the arts through a nomadic life aboard a sailing ship. Other artists emphasize their inclusive and open practices through locations such as \texttt{"earth"} for \href{https://github.com/keroserene}{Serene} who ``types code and concertos'', or \texttt{"W/O/R/L/D/W/I/D/E"} for \href{https://github.com/lee2sman}{Lee T}, a generative artist and maker.

\begin{tcolorbox}[boxrule=1pt,arc=.3em, left=2pt, right=2pt]
  \textbf{Answer to RQ2}: Generative art and creative coding rare present all around the world. We found contributors in 124 different countries, including large, active communities in Japan, Brazil, India,  Europe and North America. We also confirm the presence of creative coders in Africa, Central Asia or the Indian Ocean. This is  evidence of the large-scale adoption of code as a creative medium and an encouragement to develop these practices for sustained diversity in open-source software.
\end{tcolorbox}

\subsection*{RQ3. \rqcharacteristics}

With this question, we dive into the diversity of creative practices  among \pfive users. We use an LLM to analyze \nbpfiverqthree sketches according to the framework presented in \autoref{tab:framework}. This framework characterizes each sketch along three dimensions: \textit{Material and Processes}, \textit{Interaction}, and \textit{Sensory Outcomes}. For instance, the code in \autoref{fig:example-generative} draws shapes using random values, without requiring user input, and is classified as follows {\small \texttt{material\_processes = [synthesized\_image, randomness]}; \texttt{interaction = false}; \texttt{sensory\_outcome = [visual, static]}}.

\begin{figure}[h]
\includegraphics[width=\columnwidth]{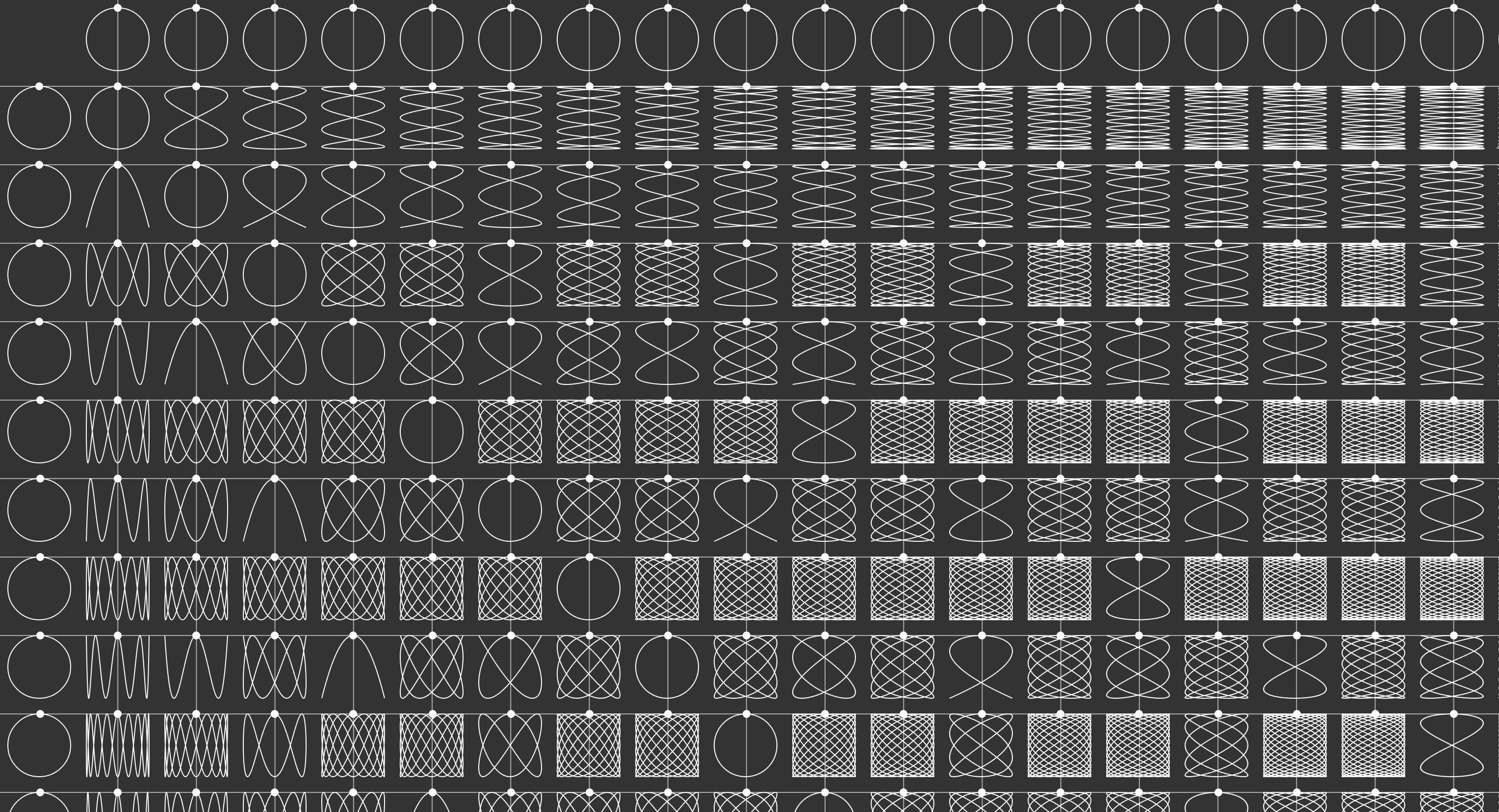}
    \caption{ Example of a static, synthesized image as generated by this \href{https://github.com/seattleacademy/website/blob/main/CodingChallenges/CC_116_Lissajous/P5/sketch.js}{sketch}.}
     \label{fig:exampleblack}
\end{figure}

We first look into the \matproc used in \pfive sketches, as summarized in  \autoref{fig:material-processes}. The left part of the figure provides the distribution of sketches that synthesize elements. The total bar gives the share of sketches that have at least one synthesis action; the next three bars distinguish between sketches labeled with \synvisual, \syntext, or \synaudio according to different types of elements, noting that one sketch can be labeled with more than one.
Almost all sketches (96.21\%) include code that relates to synthesizing new material. The vast majority (81.70\%) use code to generate visual elements. This result is expected, given that \pfive initially aimed to support visual arts. From the earliest version of the library, there have been functions to facilitate generating and placing geometric shapes on the canvas, or handling colors. 

We also observe a significant share of sketches that synthesize sound (6.61\%). Sound synthesis either relies on a third-party library such as \texttt{Tone.js}, or on a dedicated \pfive extension for sound synthesis introduced in 2024. Functions for sound synthesis include the creation of oscillators whose frequency, waveform, and amplitude can be precisely defined, and which can be further refined using envelopes, filters, and effects.  
For example, \href{https://github.com/Kyntaz/sine-rider}{sine-rider} defines 6 oscillators with different frequencies and amplitudes. 
Text synthesis (18.81\%) appears when the code prints text on the screen. As an example, some interactive artworks allow the audience to manipulate words that also contain links or typographic effects. It is important to note that the LLM labels a sketch that writes text for a button or a slider as \syntext.

Randomness is an important component in the generative art practices. As shown in the right part of \autoref{fig:material-processes}, 49.63\% of the sketches use randomness. This makes it possible to generate a different output in each execution and is a typical practice in generative art. While the concept of yielding a part of the artist's agency to the algorithm and the machine is a at the core of generative art, this can take many forms in the code, such as generating a random number or using various sorts of noise functions. In order to embrace this diversity of implementations, we instruct the LLM to account for various expressions of randomness by providing it with a variety of examples that all convey the intention of randomness.

There are also a high number of sketches that do not explicitly use randomness. In these cases, variation still occurs and is triggered by other sources, such as time, user interaction or external data. Therefore, a sketch can still be generative even though its code does not include functions that generate random numbers directly. 
In particular, the middle part of \autoref{fig:material-processes} reveals that many sketches (50.72\%) use external sources to feed the generative algorithm. Most of these sketches process images or video material as part of a work that transforms the artifact, or extracts data from it (size, density of certain pixel colors, etc.), or displays the image as part of a collage or a game.

To understand these sketches that process external artifacts, we analyze the type of artifacts they reuse in \autoref{fig:external-assets}. Overall, 34.99\% (6,256 sketches) of the dataset uses at least one external artifact. The most common type of artifact is image or video. In many cases, image artifacts are used as part of games, animations, or interactive scenes. For example, a \href{https://github.com/HaniyaHasan/project-19-}{cycling game} loads image files for characters, obstacles, and backgrounds. Many sketches also process an external artifact to extract data that is used as part of the generative process. For example, this \href{https://github.com/PrinceGrewal/website/blob/main/Tutorials/P5JS/p5.js_sound/17.8_minInput/sketch.js}{sketch} processes audio to extract its amplitude and passes it on as the size of a generated ellipse.
Data files, such as \texttt{.json} and \texttt{.csv}, also appear in many sketches. The sketches may extract data to control certain parts of the rendering, such as color, density, or duration; in other cases, data is used as textual source. We find only a few sketches that rely on external font files, which might indicate that the works that use text generally rely on the default font families provided by \pfive.

We note that the middle part of \autoref{fig:material-processes} shows that 50.72\% of sketches process image, text, or audio, while in \autoref{fig:external-assets} we find that only 34.99\% of sketches load a media artifact from the same folder as the sketch. This difference occurs when the sketch fetches the artifact to be processed dynamically from external sources. The artifact can be loaded through an HTTP request, as in this \href{https://github.com/stefiHB/NOC-S17-2-Intelligence-Learning/blob/master/week6-rnn-tensorflow/01b_rnn_flask_js/static/sketch.js}{sketch}. It can also be passed by a user, for example in a text box or through sensors, as in this \href{https://github.com/gabriel-combe/Rainbow-Code/blob/master/Tutorials/P5JS/p5.js_video/10.1_p5.js_createCapture/sketch.js}{sketch} that fetches a video feed from a webcam.

We now look at the diversity of experiences for the audience, with labels regarding \interaction and \outcome, in \autoref{fig:interaction-outcome}. The figure shows four quadrants, comparing \interactive, \noninteractive, \static, and \timebased outcomes. In each quadrant, one Venn diagram represents the distribution of sketches that have \auditory, \visual, or both outcomes.

A vast majority of the sketches produce a \timebased outcome, as observed in the two quadrants on the right of the figure. Among these, 60.52\% are also \interactive, and 39.45\% are \noninteractive. The sketches that produce \visual, \interactive, \timebased outcomes correspond to animations on the web page, games, or interactive visual artworks. Meanwhile, the \visual, \noninteractive, and \timebased sketches are autonomous animations, such as particle systems, simulations, autonomous drawing systems, and cellular automata. For example, this \href{https://github.com/hcGamechamp/hcGamechamp.github.io/blob/main/Walker%20OOP/sketch.js}{sketch} creates three walkers that autonomously move at a constant speed, following a random path. 

An \auditory outcome is necessarily \timebased, as sound has to unfold over time. Therefore, the quadrants on the right show the share of sketches that are \auditory or \audiovisual. Manual inspection reveals that many of the sketches that produce an \audiovisual outcome are web page animations or games, such as the cycling game mentioned previously. An example of non-interactive audiovisual can be found in this \href{https://github.com/seattleacademy/website/blob/main/CodingChallenges/CC_110.2_recaman_music/P5/sketch.js}{sketch} where spirals and sound unfold over time, autonomously. Purely auditory sketches are few, with only 179 sketches in total. In this group, the interactivity allows the audience to control some parameters of synthesized sounds, such as the pitch, the volume or the shape of the oscillator. 

We now look at the left quadrants of \autoref{fig:interaction-outcome}. They show that only 21.89\% of the sketches produce a \static outcome, including 14.85\% that are \noninteractive and 7.04\% that are \interactive. The former category implements an algorithm that draws an image on the screen, which does not move, as in \autoref{fig:exampleblack}. Manual inspection of the latter category of sketches reveals that they implement interactive features such as a text box or mouse capture, but these interactions do not affect the generation of the visuals.

\begin{tcolorbox}[boxrule=1pt,arc=.3em, left=2pt, right=2pt]
  \textbf{Answer to RQ3}: \pfive is a versatile library used by artists and creative coders across a wide range of practices. 
  A vast majority of the sketches produce a \visual outcome, yet almost 10\% combine \auditory and \visual outcomes. Most of the works are \timebased. Manual inspection of sampled sketches reveals that they range from abstract artworks to web animations and games. These findings are useful for educators to embrace the variety of practices in their curricula, and for researchers  to design techniques for fine-grained qualification of generative sketches.
\end{tcolorbox}

\subsection*{RQ4. \rqtop}

\begin{figure}[t]
  \centering
  \includegraphics[width=1\linewidth]{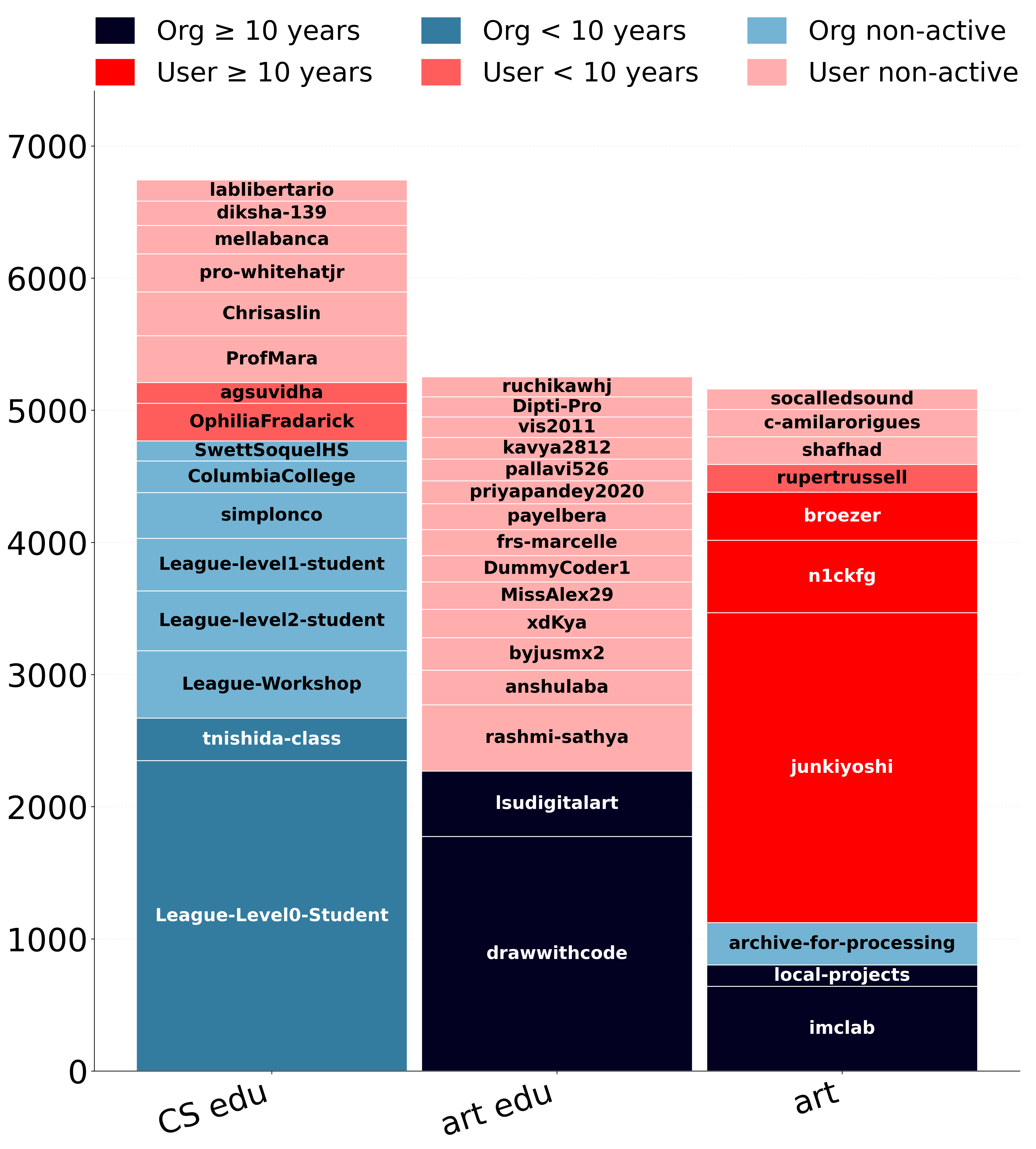}
  \caption{Distribution of the \nbreposbytopfortytwo repositories associated with the top \nbtopusers contributors in our dataset, grouped by purpose (\emph{art}, \emph{art education}, and \emph{computer science education}) and activity status (\emph{active} vs. \emph{non-active}). Each segment represents one user or organization, with height proportional to the number of repositories.}
  \label{fig:plot3}
\end{figure}

In order to address this question, we manually analyzed a subset of repositories from our dataset. We selected the \nbtopusers GitHub accounts associated with the largest number of repositories. The subset is therefore made up of \nbreposbytopfortytwo repositories. The account that has the largest number of art-related repositories belongs to the GitHub organization \texttt{League-Level0-Student}, with 2,348 repositories, followed closely by the individual account \texttt{junkiyoshi}, with 2,344 repositories. The 42nd account, owned by user \texttt{ruchikawhj}, has 152 repositories.

Our analysis reveals three main purposes that characterize how these repositories are created and used, in relation to generative art: \emph{computer science (CS) education}, \emph{art education}, and \emph{art}. 
The \emph{CS education} category includes accounts created to teach and learn about programming concepts through generative art and creative coding, and is mainly represented by educational organizations and educator accounts.
The \emph{art education} category includes accounts developed for teaching and learning contexts related to creative programming, visual design, and artistic experimentation, often by universities, art programs, or educators. 
The \emph{art} category includes repositories created primarily with an artistic point of view, either by individual artists or by art-oriented organizations such as studios or collectives. 

\autoref{fig:plot3} displays how the \nbtopusers accounts and their repositories \nbreposbytopfortytwo are distributed according to their primary purpose. In the figure, each segment represents one GitHub account, with height proportional to the number of repositories associated with that account. We use two colors and three shades within each color to distinguish between accounts belonging to individual users (orange) and organizations (blue), as well as between non-active accounts (lightest shade), accounts active for less than 10 years (medium shade), and accounts active for more than 10 years (darkest shade).

Activity status helps to distinguish between sustained and time-bounded forms of engagement with public code. The figure shows a relatively high number of non-active accounts, particularly in the two educational categories, where participation is often tied to a specific period of study, teaching, or course involvement. Although the \emph{computer science education} category contains the largest number of repositories, many of its contributors were active between 2020 and 2022, a period that coincided with the COVID-19 pandemic and widespread stay-at-home restrictions, when teaching and learning moved to online platforms. This suggests that a substantial part of these repositories was created for specific educational contexts, such as courses, assignments, or teaching materials, rather than as part of a long-term artistic practice. By contrast, the \emph{art} category shows a higher proportion of active contributors, indicating that engagement with public code for artistic purposes is more likely to persist over time. The \emph{art education} category occupies an intermediate position, combining active artistic organizations and pedagogical initiatives with contributions from non-active users. At the same time, the fact that many educational accounts are no longer active on GitHub does not necessarily mean that these practices disappeared; rather, it may indicate that they moved to other platforms or to less publicly visible forms of code sharing.

\textbf{CS education}: In this category, generative art technologies such as \processing or \pfive are mainly used as an engaging way to introduce programming concepts through visual and interactive output. In many of these cases, each student creates one repository per assignment or module within a GitHub organization, and these repositories accumulate over multiple courses and academic years. An example of an individual account in this category is \texttt{\href{https://github.com/OphiliaFradarick}{OphiliaFradarick}}, a data analyst who worked as a coding instructor for \texttt{\href{https://github.com/whitehatjr}{whitehatjr}}, an online coding school for children, between 2020 and 2023. In our top \nbtopusers, we also identify one account related to the same \texttt{whitehatjr} organization, \texttt{\href{https://github.com/pro-whitehatjr}{pro-whitehatjr}}. The repositories associated with \texttt{OphiliaFradarick} account include forks of projects from \texttt{pro-whitehatjr} as well as student assignments, illustrating how code in this category forms part of structured teaching activities. Another example of an organizational account is \texttt{\href{https://github.com/ColumbiaCollege}{ColumbiaCollege}}, which corresponds to Columbia College, a public community college in Sonora, California, USA. The repositories in this account mainly use \processing and store student assignments and midterm projects from 2019. This account is associated with 240 repositories in our dataset. 
In general, individual accounts in this category are often associated with students and freelance educators, many of whom were active between 2020 and 2022. By contrast, organizational accounts are more likely to remain active over time because they are sustained by recurring students, multiple instructors, and the ongoing reuse of GitHub as part of course administration and submission workflows.

\textbf{Art education}: This category reflects the use of creative coding in teaching and learning environments for art and design students. It includes accounts associated with courses, workshops, studio-based learning, and institutional programs, such as university initiatives, in which students experiment with code while developing both technical and creative skills. Compared with \emph{computer science education}, this category places greater emphasis on the exploration of visual forms, interaction, abstract ideas, and digital creativity, and less emphasis on programming concepts and algorithms. An example of a GitHub organization in this category is \texttt{\href{https://github.com/lsudigitalart}{lsudigitalart}}. The documentation of the repositories in this account reveals that it relates to an online course that teaches students contemporary web design and development workflows, with particular emphasis on interaction and visualization. It stores student homework and project submissions, which helps explain its large number of repositories. The account remains active and has been active since 2016, suggesting a pedagogical use of GitHub in this context. In contrast, individual accounts in this category are often associated with students. For example, \texttt{\href{https://github.com/rashmi-sathya}{rashmi-sathya}} represents a non-active student account in our top list. This account contains one repository per project, many of which use the \pfive library to create interactive pieces such as games. For instance, one repository is described as a \emph{Bouncy Ball} created using the \href{https://brm.io/matter-js/}{Matter.js} physics engine. Such examples illustrate how repositories in this category often function as records of exploratory and project-based learning in creative coding.

\textbf{Art}: By opposition to the GitHub accounts categorized as \emph{computer science education} or \emph{art education}, the accounts whose primary purpose is \emph{art} contain large numbers of repositories for a wider variety of reasons. These repositories often correspond to individual artworks, installation projects, technical experiments, reusable tools, or forks of open-source technologies relevant to artistic production. This diversity reflects the broad range of ways in which code is used as a creative medium in artistic practice. The majority of repositories in the \emph{art} category represent artworks, prototypes, or experiments conceived by artists or art-oriented organizations rather than structured educational tasks. Individual artists and organizations often create one repository per artwork, installation, or technical exploration. In some cases, contributors maintain sustained creative routines, publishing new sketches, experiments, or artworks over several years. As a result, repositories in this category function as records of artistic production, experimentation, and technical evolution over time.

An example of a GitHub organization in this category is \texttt{\href{https://github.com/imclab}{imclab}}, an art gallery based in New York City. Its repositories include both original projects and forks, suggesting a practice that combines artistic production with the adaptation and reuse of existing open-source technologies. The range of tools and languages present in this account indicates engagement with multiple technical ecosystems relevant to digital art production, including creative coding and multimedia programming environments. An example of an individual account is \texttt{\href{https://github.com/n1ckfg}{n1ckfg}}, whose profile combines artistic practice with technical research and development. This account includes personal projects, forks, and tool-oriented repositories (like AEToolbox UI panel) spanning multiple languages and domains, including animation, 3D drawing, sound-related work, and interfaces for artistic production.

To illustrate the diversity of contributors in our dataset, we examine the three largest accounts. These contributors are informative because they represent different regions, purposes, and technical ecosystems, while also showing a long-term engagement with public code related to generative art. 

\texttt{\href{https://github.com/League-Level0-Student}{League-Level0-Student}} is the largest contributor, with 2,348 repositories. This account belongs to \emph{The League of Amazing Programmers}, an educational organization based in California, USA, that teaches programming to students from 3rd to 12th grade. 
It is representative of the \emph{computer science education} category, with repositories organized around classes, levels, modules, and assignments. In many cases, each repository corresponds to the work of one student or a group of students on a specific assignment. The account also includes templates, forks, and course materials that support teaching activities. For example, in Module 1 of Level 0, students learn introductory programming concepts such as conditionals (\texttt{if/else}), variable initialization, and \texttt{for} loops, while creating simple graphical outputs with \processing. This structure explains the large number of repositories: GitHub is used here primarily to organize, submit, and store student work over time. The League is also associated with other accounts in our top \nbtopusers contributors, including \texttt{League-Workshop} ranked 6th, \texttt{League-Level2-Student} ranked 9th, and \texttt{League-Level1-Student} ranked 10th. Together, these accounts contribute 3,309 of the 6,742 repositories in the \emph{computer science education} category, representing 55\% of that subset. 

\href{https://github.com/junkiyoshi}{\texttt{junkiyoshi}} represents a different profile. This individual account belongs to NAKAUCHI Kiyoshi, a software developer based in Japan whose public GitHub activity reflects a daily engagement with generative art, particularly with \OF. Since 2016, he has created one repository per day, pushing one artwork to that repository and sharing the results on social media platforms such as Instagram or X. In our dataset, this account is associated with 2,344 repositories; as of April 16, 2026, his GitHub profile reported approximately 3.2 thousand repositories. His artworks frequently explore movement, randomness, and geometric forms such as triangles, spheres, and particle-based compositions, often rendered in either vivid color palettes or black and white. \emph{junkiyoshi} provides a clear example of long-term individual artistic engagement with code. 

\href{https://github.com/drawwithcode}{\texttt{drawwithcode}} is an organization focused on the education of art and design. It is associated with the elective course \emph{Creative Coding} at Politecnico di Milano, Italy, where code is taught as a generative tool for design students rather than a technical skill. Its repositories are primarily based on JavaScript and \pfive, and are linked to student assignments and projects developed within the course. The large number of repositories can be explained by the course structure: initially, each student team’s project was stored in a separate repository for each course edition, while more recently GitHub Classroom has been used to automatically create and manage repositories for student submissions. The account is also connected to DensityDesign, a research lab in the Design Department focused on the visual representation of complex social, organizational, and urban phenomena.

\begin{tcolorbox}[boxrule=1pt,arc=.3em, left=2pt, right=2pt]
  \textbf{Answer to RQ4}: The manual analysis of the top \nbtopusers contributors and their \nbreposbytopfortytwo source code repositories reveals that public code in generative art serves for three main purposes: computer science education, where code for the art is used as creative way of learning; art education, where teaching focuses on code as an art and design medium; and art production, where code is produced, reused and remixed to develop novel art installations. 
\end{tcolorbox}

\section{Discussion}

In this section, we discuss the implications of our study for the field of software research, and the threats to validity.

\subsection{Implication for Software Engineering Research}

\paragraph{Art is a Software Application Domain} 
When using code as a creative medium, artists and creative coders have access to the full spectrum of software technology. Consequently, new media installations, interactive artworks, and blockchain-based generative artworks are full-fledged software systems. While they share some challenges with any other software system, they also have their own, which can be further investigated by the software research community. For example, \processing and \pfive have grown from libraries into full ecosystems of third-party libraries that address specific concerns in generative art, such as handling sound or gameplay. These specialized libraries support the many practices we have observed in RQ3. Yet, the software supply chain of generative art is vast and much understudied.

In addition to third-party libraries, we have observed that generative artworks depend on many other types of artifacts, such as audiovisual assets, specific protocols to control multiple devices, or third-party APIs to fetch data that feeds the artwork. These various types of dependencies contribute to the fragility of these artworks. The software engineering research community can contribute dedicated techniques to precisely document these dependencies in order to support artists and cultural workers in the preservation of these works.

We have seen that the \swh archive plays a key role in the preservation of art-related source code. The restoration of early code-based artworks will also require specific migration techniques. For example, Adobe Flash and Java applets were popular for net.art pieces in the early 2000s, and traces of source code can be found in archives. However, it is hard to perform these pieces on modern computers or even portable devices. The software engineering research community can contribute code transformation and migration techniques that aim at restoring or facilitating access to software-based artworks.

\paragraph{Software Literacy with Creative Coding} When answering our four research questions, we have manually inspected hundreds of code repositories from our dataset. This inspection reveals that a significant share of the repositories are authored by students. These student repositories include assignments for various courses related to art and creative coding. In particular, we note the significant influence of Daniel Shiffman's \href{https://github.com/CodingTrain}{coding train} tutorials and learning material that help students learn about the basic concepts of programming, all the way to advanced physics simulations and interactive technology for immersive artworks.

These repositories are evidence of the role that creative coding plays to introduce software concepts to diverse audiences. Our work can inspire software educators to explore art as an entry point to software development, as well as a way to demystify software technology and increase software literacy in diverse communities.

\subsection{Threats to Validity}

Here we discuss the threats to the validity of our study and how we have mitigated them. 

The first threat relates to our selection process, which reveals a worldwide phenomenon, yet through a limited lens. Our set of \nborigins repositories might well just be a drop in the ocean of humanity's code for the arts. We have selected code repositories based on the presence of certain file-level signals. Consequently, we omit many pieces that do not meet our selection criteria, including: art-related public repositories that call generative art libraries in the file contents or those that do not use generative art libraries at all, such as the CryptoPunks~\cite{larva17} \footnote{\url{https://github.com/larvalabs/cryptopunks}}; art projects that use other libraries like Blender or a game engine, such as Unity; and all the historical code-based artworks for which source code is not available online, such as pieces from the demoscene or the code that Vera Molnar wrote by hand in her diaries \cite{molnardiaries}. Also, we mine from \swh, who currently does not archive public code that is published on art-specific platforms such as OpenProcessing\footnote{\url{https://openprocessing.org/}} or the code of NFTs that is available on IPFS~\cite{RCSfxhash}. Last but not least, we can only analyze code repositories that are shared publicly. We do not have access to all the code that is developed within art and design studios, as well as in creative companies that do not share their code. Despite these limitations, our selection criteria, which we motivate in \autoref{sec:corpus}, allow us to collect a large, diverse, and historically relevant set of code repositories that have an artistic intention as demonstrated throughout this work.

The second threat relates to the correctness of the code we use to analyze these source code repositories. We identify two threats here: one related to the quality of the code we write; the other related to the quality of the data and third-party tools we use. This latter point is particularly relevant to RQ2, where we analyze locations derived from arbitrary strings entered by users and rely on a geocoding service provided by Geoapify to resolve them. We mitigate these threats through several rounds of manual testing and analysis of data samples to clean the data and ensure the quality of the results. This results in a conservative lower bound on the number of locations we analyze and calls for finer-grained geospatial analysis in the future.

\section{Conclusion}

In this paper, we have explored the public source code that thousands of artists, creative coders, students, and educators have contributed across the world. 
The study provides the following insights: while most of the code is shared on GitHub, various maker and artist communities  share on alternative code hosting platforms;  active artists and communities are present all around the world;  practitioners use \pfive to develop a wide variety of practices, from randomized abstract images to interactive \audiovisual artworks or games;  art-related code serves various purposes that span programming and art education to professional art and design practices.

Generative art might be as ancient as art itself~\cite{timeline26}, and as soon as artists have had access to computers, they have explored code as a medium to build generative systems. 
This first large-scale empirical study of public code repositories related to art and creative coding provides novel insights for artists and software developers who develop new libraries and tools for digital art;  for software and art educators who aim at introducing programming concepts through creative technology; for software and art professionals who care about the preservation of our digital cultural heritage. 
Our results open several threads of future work.
First, we wish to investigate to what extent these projects still execute and what the strategies are to restore the broken projects. Second, we aim to model the various types of software, data, and network dependencies that underlie these artworks, in order to support conservation and restoration practices for generative art.

\section*{Data Availability}

The complete set of art-related code repositories is available on \href{https://zenodo.org/records/20184094}{Zenodo}, and all the scripts and intermediate datasets to answer the research questions are available as part of our reproduction package on GitHub: \href{https://github.com/sparkrew/art-in-humanitys-code}{art-in-humanitys-code}, archived on Software Heritage with SWHID \href{https://archive.softwareheritage.org/swh:1:rev:1b4edb9beeee3a15687273e70a5500f35a135e8a}{swh:1:rev:1b4edb9beeee3a15687273e70a5500f35a135e8a}.

\section*{Acknowledgements}
We thank Geoapify for granting free access to their infrastructure for research purposes. We thank \href{https://github.com/anacat}{anacat from Braga, Portugal} and the thousands of generative artists who have inspired this work. This research was enabled in part by support provided by Calcul Québec (calculquebec.ca/) and the Digital Research Alliance of Canada (alliancecan.ca). This research was undertaken thanks to funding from the Canada Research Chairs Program, FRQ, NSERC and from SSHRC. This work was made possible by Software Heritage, the universal source code archive: \url{https://www.softwareheritage.org}.

\bibliographystyle{IEEEtran}
\bibliography{biblio}

\begin{thebibliography}{10}
\providecommand{\url}[1]{#1}
\csname url@samestyle\endcsname
\providecommand{\newblock}{\relax}
\providecommand{\bibinfo}[2]{#2}
\providecommand{\BIBentrySTDinterwordspacing}{\spaceskip=0pt\relax}
\providecommand{\BIBentryALTinterwordstretchfactor}{4}
\providecommand{\BIBentryALTinterwordspacing}{\spaceskip=\fontdimen2\font plus
\BIBentryALTinterwordstretchfactor\fontdimen3\font minus \fontdimen4\font\relax}
\providecommand{\BIBforeignlanguage}[2]{{%
\expandafter\ifx\csname l@#1\endcsname\relax
\typeout{** WARNING: IEEEtran.bst: No hyphenation pattern has been}%
\typeout{** loaded for the language `#1'. Using the pattern for}%
\typeout{** the default language instead.}%
\else
\language=\csname l@#1\endcsname
\fi
#2}}
\providecommand{\BIBdecl}{\relax}
\BIBdecl

\bibitem{paul2023digital}
C.~Paul, \emph{Digital art}.\hskip 1em plus 0.5em minus 0.4em\relax Thames \& Hudson, 2023.

\bibitem{cotton_radical_2024}
M.~Cotton, T.~Rivers~Ryan, and M.~Rosen, \emph{{Radical Software: Women, Art \& Computing, 1960-1991}}.\hskip 1em plus 0.5em minus 0.4em\relax Verlag der Buchhandlung Walther und Franz König, 2024.

\bibitem{molnar1975toward}
V.~Molnar, ``{Toward Aesthetic Guidelines for Paintings with the Aid of a Computer},'' \emph{Leonardo}, vol.~8, pp. 185--189, 1975.

\bibitem{sykora1970computer}
Z.~S{\`y}kora and J.~Bla{\v{z}}ek, ``Computer-aided multi-element geometrical abstract paintings,'' \emph{Leonardo}, vol.~3, no.~4, pp. 409--413, 1970.

\bibitem{cornock1973creative}
S.~Cornock and E.~Edmonds, ``The creative process where the artist is amplified or superseded by the computer,'' \emph{Leonardo}, vol.~6, no.~1, pp. 11--16, 1973.

\bibitem{mccormack2001}
J.~McCormack, A.~Dorin \emph{et~al.}, ``Art, emergence and the computational sublime,'' in \emph{Proceedings of Second Iteration: A Conference on Generative Systems in the Electronic Arts. Melbourne: CEMA}, 2001, pp. 67--81.

\bibitem{puckette1996pure}
M.~Puckette \emph{et~al.}, ``{Pure Data: Another Integrated Computer Music Environment},'' \emph{Proceedings of the second intercollege computer music concerts}, pp. 37--41, 1996.

\bibitem{mccartney2002rethinking}
J.~McCartney, ``{Rethinking the Computer Music Language: SuperCollider},'' \emph{Computer Music Journal}, vol.~26, no.~4, pp. 61--68, 2002.

\bibitem{maeda2001design}
J.~Maeda, \emph{Design by Numbers}.\hskip 1em plus 0.5em minus 0.4em\relax MIT press, 2001.

\bibitem{reas2007processing}
C.~Reas and B.~Fry, \emph{Processing: a Programming Handbook for Visual Designers and Artists}.\hskip 1em plus 0.5em minus 0.4em\relax {MIT Press}, 2007, vol. 6812.

\bibitem{noble2009programming}
J.~Noble, \emph{Programming Interactivity: A Designer's Guide to Processing, Arduino, and Openframeworks}.\hskip 1em plus 0.5em minus 0.4em\relax O'Reilly Media, Inc., 2009.

\bibitem{mccarthy2015getting}
L.~McCarthy, C.~Reas, and B.~Fry, \emph{Getting Started with p5.js: Making Interactive Graphics in JavaScript and Processing}.\hskip 1em plus 0.5em minus 0.4em\relax Maker Media, Inc., 2015.

\bibitem{boden2009generative}
M.~A. Boden and E.~A. Edmonds, ``{What is Generative Art?}'' \emph{Digital Creativity}, vol.~20, no. 1-2, pp. 21--46, 2009.

\bibitem{peppler2005creative}
K.~Peppler and Y.~Kafai, ``{Creative Coding: Programming for Personal Expression},'' in \emph{Proceedings of the International Conference on Computer Supported Collaborative Learning (CSCL)}, 2009.

\bibitem{ipres-2017-software-heritage}
R.~Di~Cosmo and S.~Zacchiroli, ``Software heritage: Why and how to preserve software source code,'' in \emph{Proceedings of the International Conference on Digital Preservation (iPRES)}, 2017.

\bibitem{islam2025trigraph}
A.~Islam and A.~Hindle, ``{TriGraph: A Probabilistic Subgraph-Based Model for Visual Code Completion in Pure Data},'' in \emph{Proceedings of the International Conference on Mining Software Repositories (MSR)}, 2025, pp. 737--749.

\bibitem{fredericks2023generative}
E.~M. Fredericks, A.~C. Diller, and J.~M. Moore, ``Generative art via grammatical evolution,'' in \emph{Proceedings of the International Workshop on Genetic Improvement (GI)}, 2023, pp. 1--8.

\bibitem{angert2023spellburst}
T.~Angert, M.~Suzara, J.~Han, C.~Pondoc, and H.~Subramonyam, ``{Spellburst: A node-based interface for exploratory creative coding with natural language prompts},'' in \emph{{Proceedings of the Symposium on User Interface Software and Technology (UIST)}}, 2023, pp. 1--22.

\bibitem{ossta18}
L.~L. McCarthy, T.~Hughes, and G.~Levin, ``{Open Source Software Toolkits for the Arts (OSSTA): a Convening},'' \url{https://github.com/CreativeInquiry/OSSTA-Report}, 2021, accessed: 2026-06-25.

\bibitem{trifonova2008software}
A.~Trifonova, L.~Jaccheri, and K.~Bergaust, ``Software engineering issues in interactive installation art,'' \emph{International Journal of Arts and Technology}, vol.~1, no.~1, pp. 43--65, 2008.

\bibitem{wu2024survey}
Z.-W. Wu, H.~Qu, and K.~Zhang, ``A survey of recent practice of artificial life in visual art,'' \emph{Artificial Life}, vol.~30, no.~1, pp. 106--135, 2024.

\bibitem{verano2023art}
M.~Verano~Merino and J.~P. S{\'a}enz, ``The art of creating code-based artworks,'' in \emph{Proceedings of the Extended Abstracts of the Conference on Human Factors in Computing Systems (CHI)}, 2023, pp. 1--7.

\bibitem{subbaraman2023forking}
B.~Subbaraman, S.~Shim, and N.~Peek, ``{Forking a sketch: How the Openprocessing Community uses remixing to Collect, Annotate, Tune, and Extend Creative Code},'' in \emph{Proceedings of the Designing Interactive Systems Conference (DIS)}, 2023, pp. 326--342.

\bibitem{baudry25}
B.~Baudry, E.~N. Gustafsson, R.~Kaufman, and M.~Euler, ``Myriad people open source software for new media arts,'' in \emph{Proceedings of International Conference on Mining Software Repositories (MSR)}, 2025.

\bibitem{IslamEH24}
A.~Islam, K.~Eng, and A.~Hindle, ``{Opening the Valve on Pure-Data: Usage Patterns and Programming Practices of a Data-Flow Based Visual Programming Language},'' in \emph{{Proceedings of the International Conference on Mining Software Repositories (MSR)}}, 2024, pp. 492--497.

\bibitem{BogdanMM25}
A.~Bogdan, M.~V. Merino, and I.~Malavolta, ``{The Ecosystem of Open-Source Music Production Software - {A} Mining Study on the Development Practices of {VST} Plugins on GitHub},'' in \emph{{Proceedings of the International Conference on Mining Software Repositories (MSR)}}, 2025, pp. 513--525.

\bibitem{payne2025exploring}
W.~Payne, M.~Kaney, I.~Rodrigues, and A.~Hurst, ``{Exploring Technical and Creative Posts in Online Live Coding Communities: An Analysis of Tidal Club and in\_thread},'' in \emph{{Companion Proceedings of the ACM International Conference on Supporting Group Work (GROUP)}}, 2025, pp. 15--21.

\bibitem{milewicz2019characterizing}
R.~Milewicz, G.~Pinto, and P.~Rodeghero, ``Characterizing the roles of contributors in open-source scientific software projects,'' in \emph{Proceedings of the International Conference on Mining Software Repositories (MSR)}, 2019, pp. 421--432.

\bibitem{trinkenreich2024unraveling}
B.~Trinkenreich, M.~A. Gerosa, and I.~Steinmacher, ``Unraveling the drivers of sense of belonging in software delivery teams: Insights from a large-scale survey,'' in \emph{Proceedings of the International Conference on Software Engineering (ICSE)}, 2024, pp. 1--12.

\bibitem{hyrynsalmi2025making}
S.~M. Hyrynsalmi, S.~Baltes, C.~Brown, R.~Prikladnicki, G.~Rodriguez-Perez, A.~Serebrenik, J.~Simmonds, B.~Trinkenreich, Y.~Wang, and G.~Liebel, ``Making software development more diverse and inclusive: Key themes, challenges, and future directions,'' \emph{ACM Transactions on Software Engineering and Methodology}, vol.~34, no.~5, pp. 1--23, 2025.

\bibitem{vasilescu2015gender}
B.~Vasilescu, D.~Posnett, B.~Ray, M.~G. van~den Brand, A.~Serebrenik, P.~Devanbu, and V.~Filkov, ``Gender and tenure diversity in github teams,'' in \emph{Proceedings of the Conference on Human Factors in Computing Systems (CHI)}, 2015, pp. 3789--3798.

\bibitem{fredericks2024generativegi}
E.~M. Fredericks, J.~M. Moore, and A.~C. Diller, ``Generativegi: creating generative art with genetic improvement,'' \emph{Automated Software Engineering}, vol.~31, no.~1, p.~23, 2024.

\bibitem{li2020supporting}
J.~Li, J.~Brandt, R.~Mech, M.~Agrawala, and J.~Jacobs, ``Supporting visual artists in programming through direct inspection and control of program execution,'' in \emph{Proceedings of the Conference on Human Factors in Computing Systems (CHI)}, 2020, pp. 1--12.

\bibitem{sacadura2026}
R.~Sacadura, L.~Goncalo, T.~Martins, and P.~Machado, ``{ Addressing the Combinatorial Explosion of Code-Based Visual Artifacts },'' \emph{IEEE Computer}, vol.~59, no.~09, pp. 69--76, Sep. 2026.

\bibitem{li2021we}
J.~Li, S.~Hashim, and J.~Jacobs, ``What we can learn from visual artists about software development,'' in \emph{Proceedings of the Conference on Human Factors in Computing Systems (CHI)}, 2021, pp. 1--14.

\bibitem{ossart23}
``{OSACC 2023 Report},'' \url{https://opensourceart.cc/osacc-2023-report/}, 2023, accessed: 2026-06-25.

\bibitem{coleman2026}
C.~Coleman, ``{ Cultivating the Code: Community, Care, and the Clinic for Open Source Arts },'' \emph{IEEE Computer}, vol.~59, no.~09, pp. 27--36, Sep. 2026.

\bibitem{sandberg2019creative}
E.~Sandberg, ``{Creative Coding on the web in p5. js: a Library where JavaScript Meets Processing},'' Bachelor's Thesis, Blekinge Institute of Technology, Sweden, 2019.

\bibitem{woo2022problem}
K.~Woo and G.~Falloon, ``Problem solved, but how? an exploratory study into students’ problem solving processes in creative coding tasks,'' \emph{Thinking Skills and Creativity}, vol.~46, p. 101193, 2022.

\bibitem{greenberg2012creative}
I.~Greenberg, D.~Kumar, and D.~Xu, ``Creative coding and visual portfolios for cs1,'' in \emph{Proceedings of the Technical Symposium on Computer Science Education}, 2012, pp. 247--252.

\bibitem{levin2021code}
G.~Levin and T.~Brain, \emph{Code as Creative Medium: a Handbook for Computational Art and Design}.\hskip 1em plus 0.5em minus 0.4em\relax MIT Press, 2021.

\bibitem{sykes2023seeing}
J.~Sykes, M.~Grierson, R.~Fiebrink \emph{et~al.}, ``Seeing programming seeing: Exploring the pedagogical values of functional errors in creative coding,'' in \emph{Proceedings of the Conference on Computation, Communication, Aesthetics \& X}, 2023.

\bibitem{mcnutt2025slowness}
A.~M. McNutt, S.~Cohen, and R.~Chugh, ``{Slowness, Politics, and Joy: Values That Guide Technology Choices in Creative Coding Classrooms},'' in \emph{{Proceedings of the Conference on Human Factors in Computing Systems (CHI)}}, 2025, pp. 1--16.

\bibitem{knochel2015if}
A.~D. Knochel and R.~M. Patton, ``If art education then critical digital making: Computational thinking and creative code,'' \emph{Studies in Art Education}, vol.~57, no.~1, pp. 21--38, 2015.

\bibitem{dufva2018art}
T.~Dufva, ``Art education in the post-digital era-experiential construction of knowledge through creative coding,'' Ph.D. dissertation, Aalto University, 2018.

\bibitem{pietri-2020-swh-graph-dataset}
A.~Pietri, D.~Spinellis, and S.~Zacchiroli, ``The software heritage graph dataset: Large-scale analysis of public software development history,'' in \emph{Proceedings of the International Conference on Mining Software Repositories (MSR)}, 2020, pp. 1--5.

\bibitem{boldi-2020-swh-graph-compression}
P.~Boldi, A.~Pietri, S.~Vigna, and S.~Zacchiroli, ``{Ultra-Large-Scale Repository Analysis via Graph Compression},'' in \emph{Proceedings of the International Conference on Software Analysis, Evolution and Reengineering (SANER)}, 2020, pp. 184--194.

\bibitem{fontana-2024-webgraph-rs}
T.~Fontana, S.~Vigna, and S.~Zacchiroli, ``{WebGraph: The Next Generation (Is in Rust)},'' in \emph{Proceedings of the Web Conference (WWW)}, 2024, pp. 686--689.

\bibitem{brehm2017polyhedral}
U.~Brehm and E.~Schulte, ``Polyhedral maps,'' in \emph{Handbook of discrete and computational geometry}.\hskip 1em plus 0.5em minus 0.4em\relax Chapman and Hall/CRC, 2017, pp. 533--548.

\bibitem{dorin2012}
A.~Dorin, J.~McCabe, J.~McCormack, G.~Monro, and M.~Whitelaw, ``{A Framework for Understanding Generative Art},'' \emph{Digital Creativity}, vol.~23, no. 3-4, pp. 239--259, 2012.

\bibitem{frohnert2017time}
C.~Frohnert and H.~Roemich, ``Time-based media art conservation education program at nyu: Concept and perspectives,'' \emph{The Electronic Media Review}, vol.~5, 2017.

\bibitem{qwen3technicalreport}
Q.~Team, ``Qwen3 technical report,'' \emph{arXiv}, 2025.

\bibitem{larva17}
M.~Hall and J.~Watkinson, ``{CryptoPunks},'' \url{https://larvalabs.com/cryptopunks}, 2017, accessed: 2026-06-25.

\bibitem{molnardiaries}
V.~Molnar, \emph{Journaux Intimes (1976 -2020)}.\hskip 1em plus 0.5em minus 0.4em\relax Oniris, 2023.

\bibitem{RCSfxhash}
``{fxhash | The Artists},'' \url{https://www.rightclicksave.com/article/fxhash-the-artists}, accessed: 2026-05-21.

\bibitem{timeline26}
P.~Bauman, ``{Generative Art Timeline},'' \url{https://timeline.lerandom.art/}, 2026, accessed: 2026-06-25.

\end{thebibliography}

\begin{wrapfigure}{l}{0.5\columnwidth}
\includegraphics[width=0.48\columnwidth]{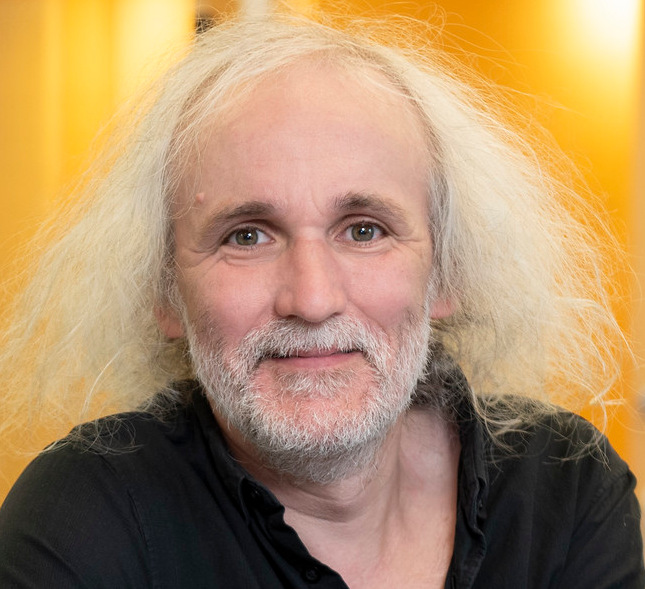}
\end{wrapfigure}
\textbf{Benoit Baudry} is a Professor in Software Engineering at the Université de Montréal. His research focuses on automated software engineering, software diversity and software testing. He favors exploring code execution over code on disk. Prior to joining Université de Montréal, he was a research scientist at INRIA (France) from 2004 to 2017, and a Professor at the KTH Royal Institute of Technology (Sweden) from 2017 to 2023.

\begin{wrapfigure}{l}{0.5\columnwidth}
\includegraphics[width=0.48\columnwidth]{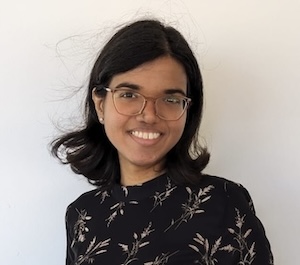}
\end{wrapfigure}\textbf{Yogya Gamage} is a PhD student at Université de Montréal, Montréal. Her research interests include software supply chain security, software testing, and software reliability. Yogya Gamage received her bachelor’s degree in computer science and engineering from the University of Moratuwa, Sri Lanka. \\

\begin{wrapfigure}{l}{0.5\columnwidth}
\includegraphics[width=0.48\columnwidth]{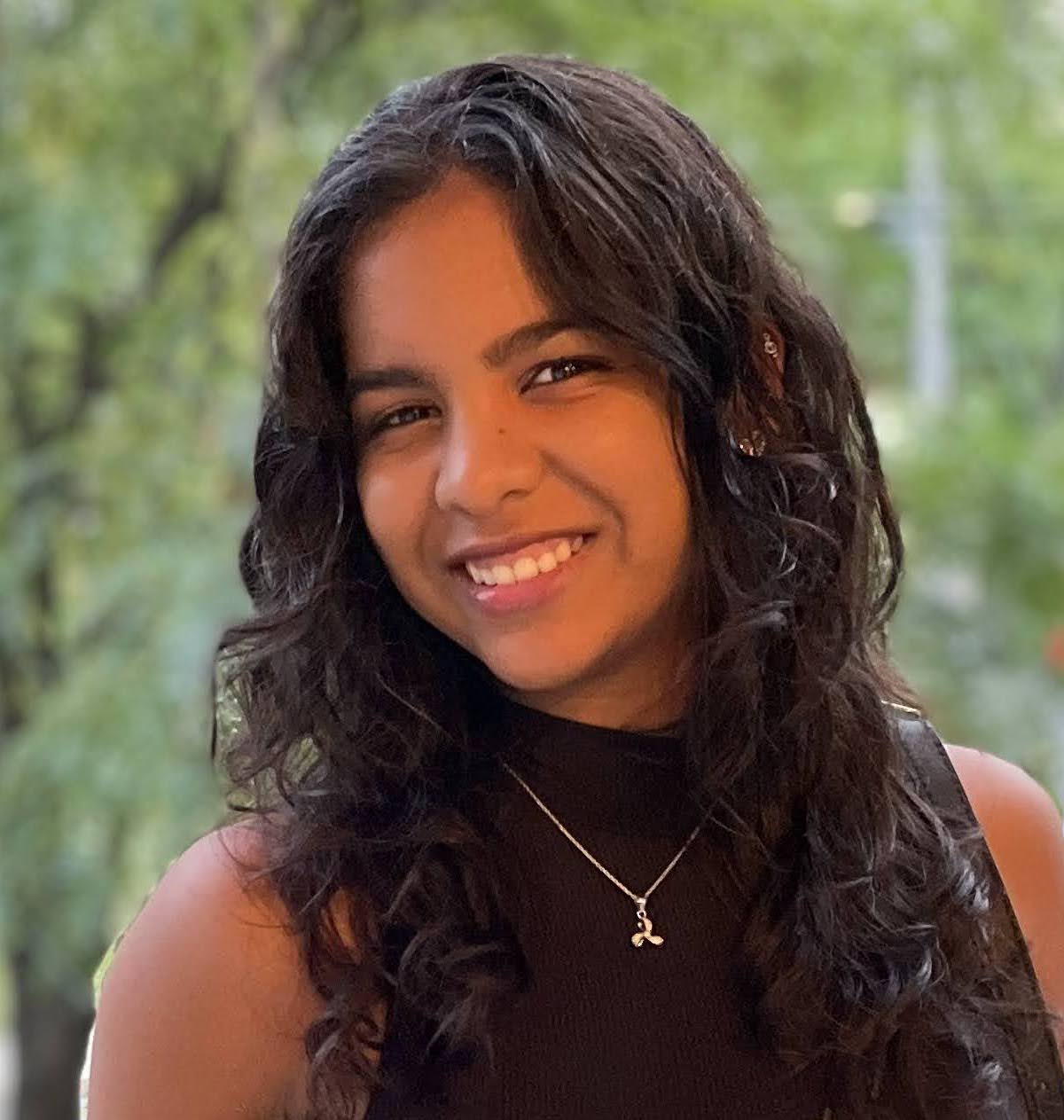}
\end{wrapfigure}
\textbf{Nadia Gonzalez Fernandez} is a Master's student in Computer Science at Université de Montréal. Her research focuses on software diversity, software supply chain security and code analysis. Nadia previously received her bachelor's degree in Computer Science from Universidad de La Habana, Cuba. \\

\begin{wrapfigure}{l}{0.5\columnwidth}
\includegraphics[width=0.48\columnwidth]{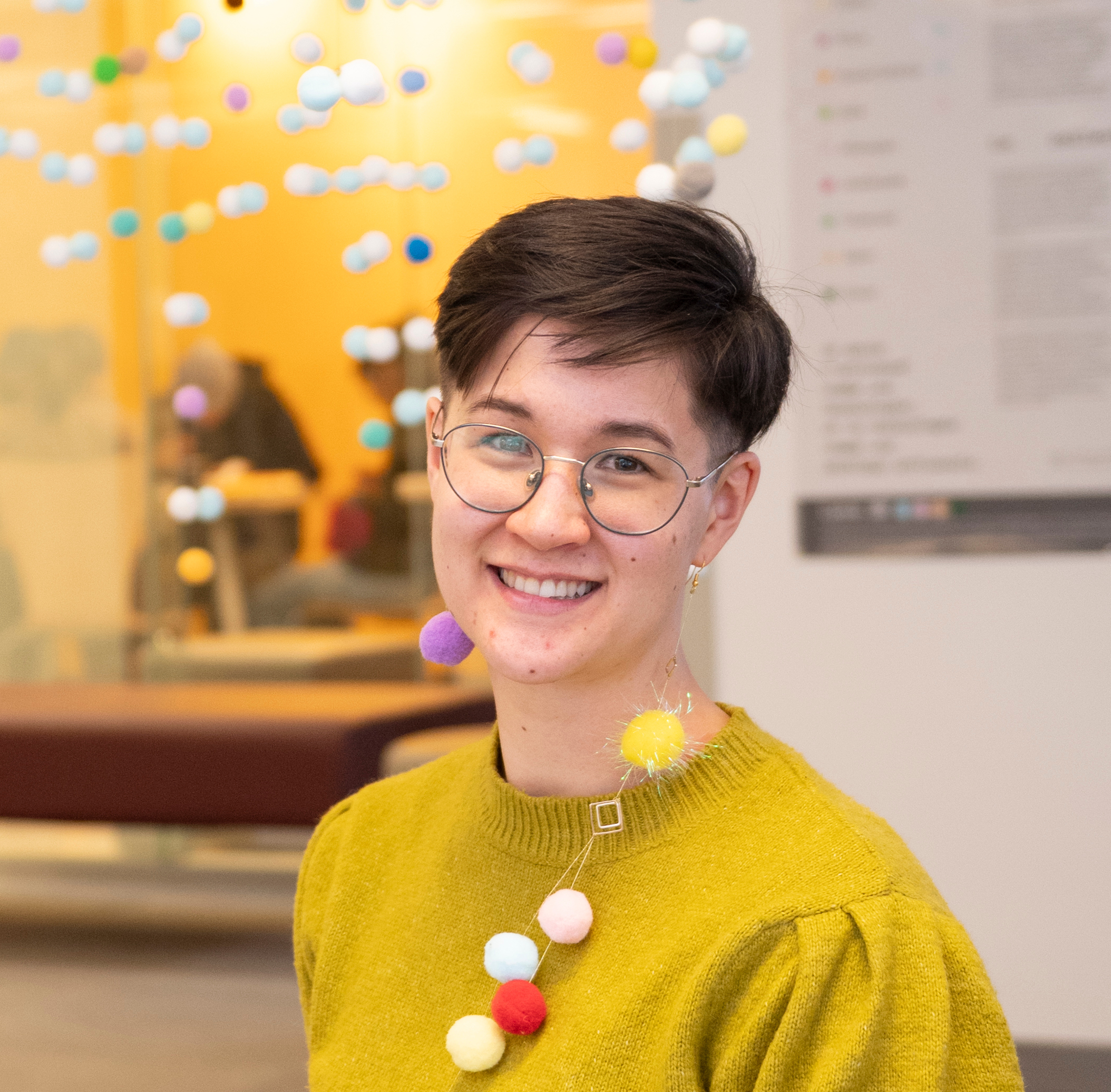}
\end{wrapfigure}
\textbf{Lena MK} is a PhD candidate at Université de Montréal. Between textile and alogorithmic practices, her research-creation spans the fields of cartography, visualization, and physicalization of cultural data. She is the founder and technical director of Maison MONA.\\

\begin{wrapfigure}{l}{0.48\columnwidth}
\includegraphics[width=0.45\columnwidth]{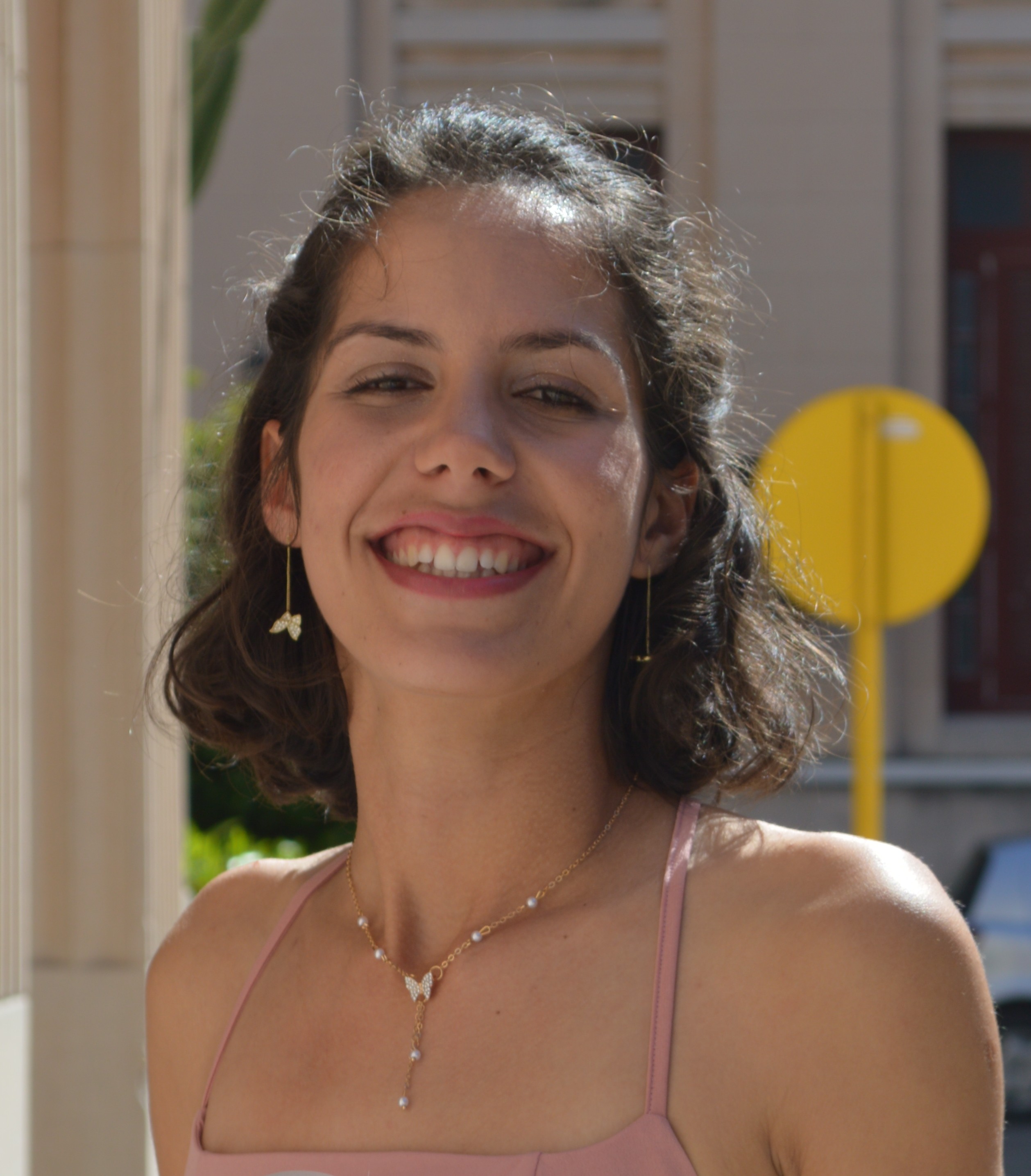}
\end{wrapfigure}\textbf{Roxana Pena Mendieta} is a master's student in Computer Science at Université de Montréal. Her research interests are focused on preservation of digital artworks, static code analysis and maintenance. Roxana received her bachelor's degree in Computer Science from Universidad de La Habana, Cuba.\\

\newpage

\begin{wrapfigure}{l}{0.48\columnwidth}
\includegraphics[width=0.45\columnwidth]{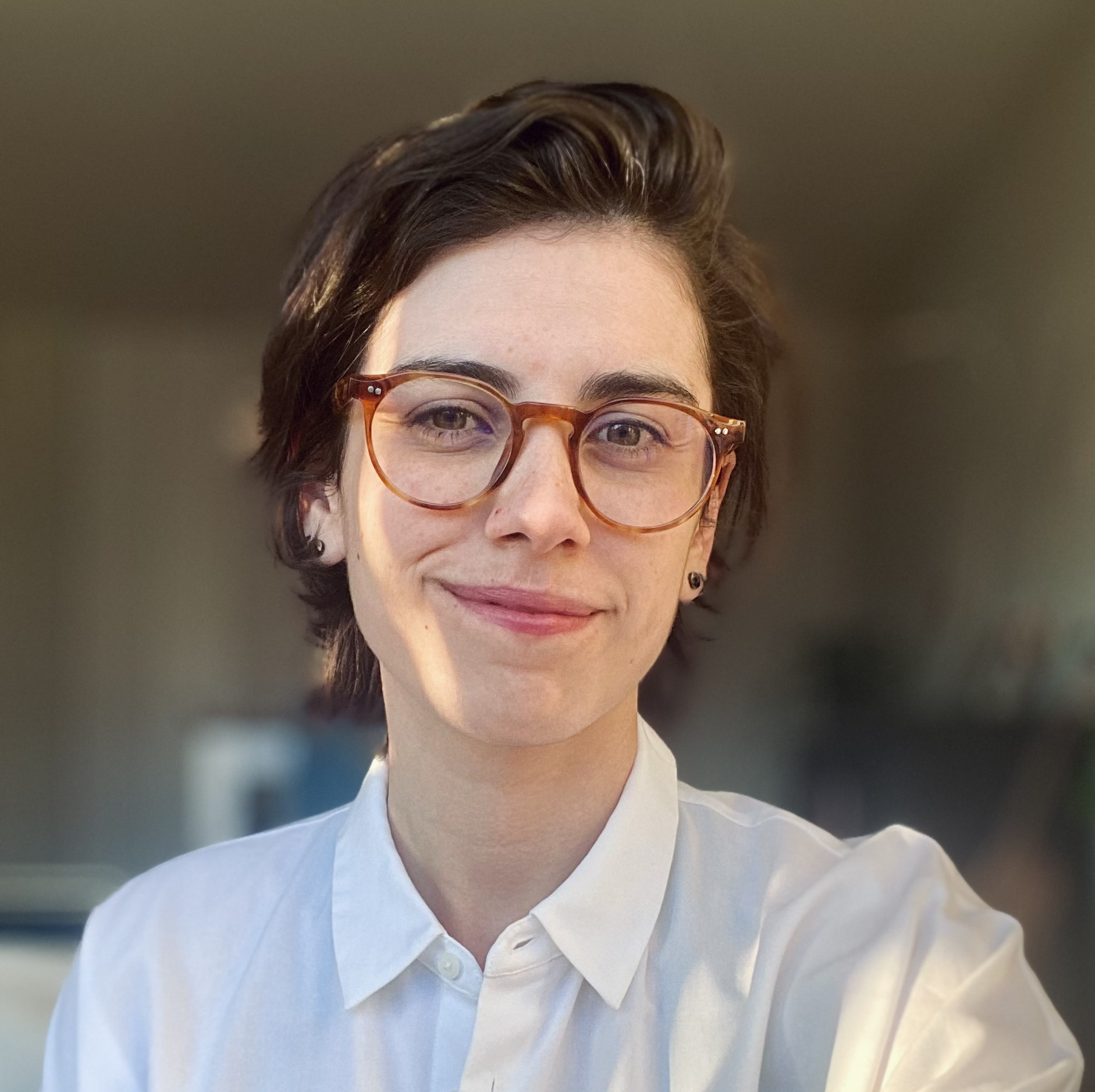}
\end{wrapfigure}\textbf{Rafaela Souza Pinter} is a Machine Learning Programmer at Ubisoft, where she develops, maintains, and scales ML systems. She holds an M.S. in Computer Science from Université de Montréal, with studies focused on deep learning, natural language processing, and probabilistic modeling. Her interests include MLOps and applying machine learning in both research and production.

\begin{wrapfigure}{l}{0.42\columnwidth}
\includegraphics[width=0.4\columnwidth]{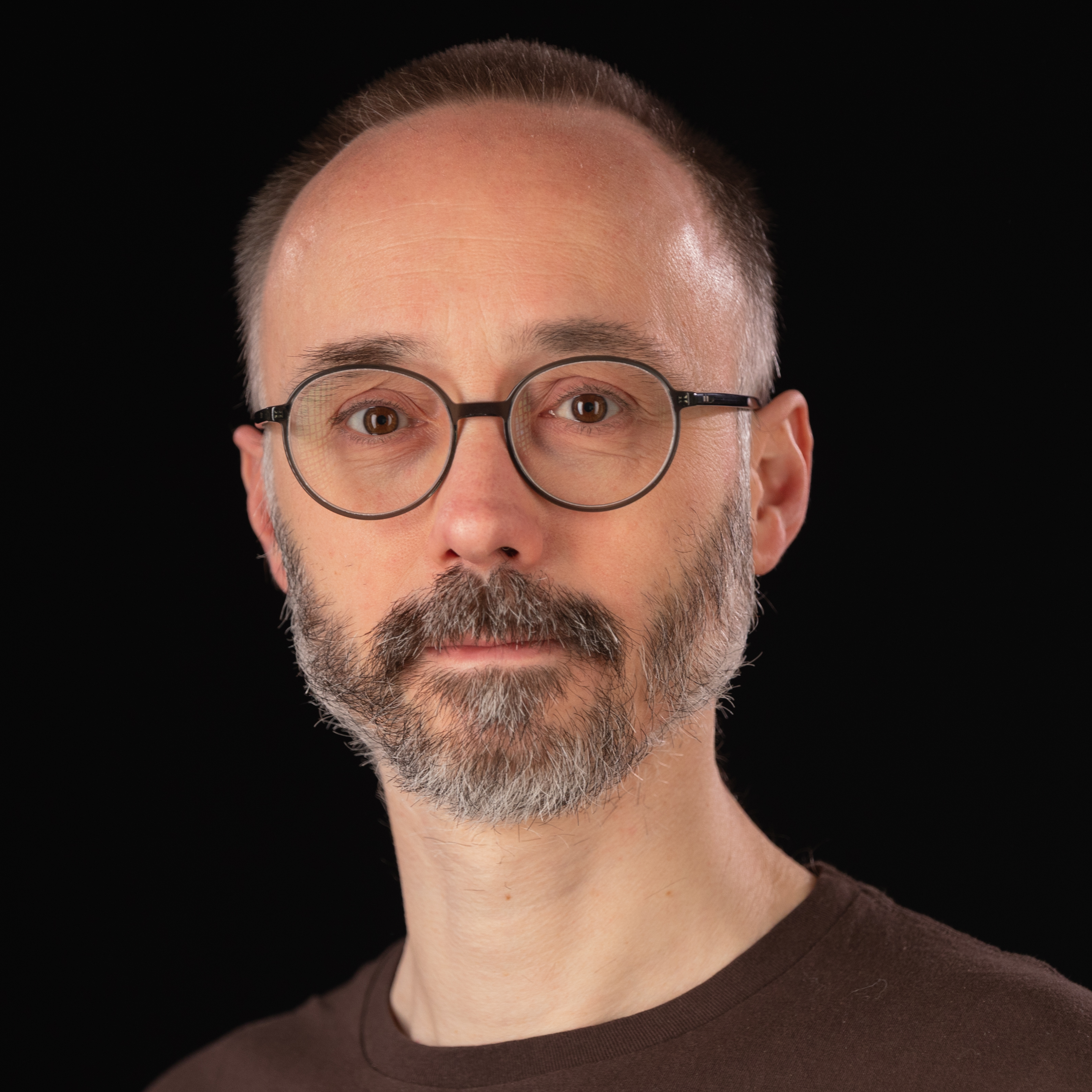}
\end{wrapfigure}\textbf{Stefano Zacchiroli} is full professor of computer science at Télécom
Paris, Polytechnic Institute of Paris. His current research interests
span digital commons, open source software engineering, computer
security, and the software supply chain. He is co-founder and Chief
Scientific Officer of Software Heritage, the largest public archive of
software source code. He is a former 3-times Debian project leader and a
former board director of the Open Source Initiative (OSI). He is
recipient of several awards, including: the 2026 European Open Source
Award, the 2022 Google Award for Inclusion Research, and the 2015
O’Reilly Open Source Award.

\end{document}